\documentclass[usenatbib]{mn2e}
\usepackage[pdftex]{graphicx}

\def\CN2{\mbox{$C_N^2  $}}

\def\CT2{\mbox{$C_T^2 \ $}}

\def\sigmal2{\mbox{$\sigma ^{2}_{I} \ $}}

\title[Re-calibrated Generalized-Scidar measurements at Cerro Paranal]{Re-calibrated Generalized-Scidar measurements at Cerro Paranal (VLT's site)}
\author[E. Masciadri et al.]{E. Masciadri$^{1}$\thanks{E-mail:
     masciadri@arcetri.astro.it}, F. Lascaux$^1$, J.J. Fuensalida$^{2,3}$, G. Lombardi$^{4}$, H. V\'azquez-Rami\'o$^{3}$\\ $^1$INAF Osservatorio Astrofisico
di Arcetri, Largo Enrico Fermi 5, I-501 25 Florence, Italy\\
$^2$Dep. Astrofisica, Universidad La Laguna, C/Astrof\'isico Francisco S\'anchez, E-38205 Tenerife, Spain\\
$^3$Instituto de Astrof\'isica de Canarias, C/Via Lactea S/N, E-38305-La Laguna, Tenerife, Spain\\
$^4$European Southern Observatory, La Silla Paranal Observatory, Av. Alonso de Cordoba, 3107 - Santiago 19, Chile}
\begin{document}
\label{firstpage}
\date{Accepted 2011 ??? ??, Received 2011 ??? ??; in original form
2011??? ??}  
\pagerange{\pageref{firstpage}--\pageref{lastpage}}
\pubyear{2011}

\maketitle

\begin{abstract}

Generalized Scidar (GS) measurements taken at the Paranal Observatory in November/December 2007 in the context of a site qualification for the future European Extremely Large Telescope E-ELT are re-calibrated to overcome the bias induced on the \CN2 profiles by a not correct normalization of the autocorrelation of the scintillation maps that has been recently identified in the GS technique. A complete analysis of the GS corrected measurements as well as of the corrected errors is performed statistically as well as on individual nights and for each time during all nights. The relative errors of the \CN2 profiles can reach up to 60$\%$ in some narrow temporal windows and some vertical slabs, the total seeing up to 12$\%$ and the total integrated turbulence J up to 21$\%$. However, the statistic analysis tells us that the absolute errors of the median values of the total seeing is 0.06 arcsec (relative error 5.6$\%$), for the boundary seeing 0.05 arcsec (relative error 5.6$\%$) and for the seeing in the free atmosphere 0.04 arcsec (relative error 9$\%$). We find that, in spite of the fact that the relative error increases with the height, the boundary and the free atmosphere seeing contribute in an equivalent way to the error on the total seeing in absolute terms. Besides, we find that there are no correlations between the relative errors and the value of the correspondent seeing. The absolute error of the median value of the isoplanatic angle is 0.13 arcsec (relative error 6.9$\%$). 

\end{abstract}

\section{Introduction}
\label{intr}
In November/December 2007 a site testing campaign (hereafter PAR2007) has been organized at Cerro Paranal by ESO with the collaboration of several international teams \citep{dali2010}. A set of different instruments run simultaneously for a number of nights variable between 10 and 20 nights with the aim to quantify different parameters characterizing the optical turbulence (the seeing $\varepsilon$, the isoplanatic angle $\theta_{0}$, the wavefront coherence time $\tau_{0}$, the spatial coherence outer scale $\mathcal L_{0}$, the vertical profile of the spatial coherence outer scale $\mathcal L_{0}$$(h)$,....). The instruments employed were: a DIMM (Differential Image Motion Monitor), a GSM (Generalized Seeing Monitor), a MASS (Multi Aperture Scintillation Sensor), a LuSci (Lunar Scintillometer), a MOSP (Monitor of Outer Scale Profiler) and a Generalized Scidar (GS). Some of these instruments (the DIMM and GSM) provide integrated values of the turbulence energy ($\varepsilon$, $\theta_{0}$, $\tau_{0}$, etc.) developed above the ground up to 20-25~km. Some of the instruments (the GS, the MASS, LuSci)  are vertical profilers of the turbulent energy (\CN2 profiles) or other astro-climatic parameter such as the spatial coherence outer scale (MOSP). One of the main goals of the PAR2007 site testing campaign was the inter-comparison of measurements that is a crucial passage not only for the characterization of a site, but, even more, for a characterization of the accuracy, uncertainty and dispersion of each instrument with respect to others. \cite{dali2010} presented a summary of the analysis of all the measurements retrieved from all these instruments from 17 December up to 26 December 2007. 

\begin{table*}
{\begin{tabular}{@{}cccccc@{}}
\hline
Name & $\alpha_{J2000}$ & $\delta_{J2000}$ & m$_{1}$ & $\Delta$m & $\theta$ \\
    & & &         (mag) & (mag) & (arcsec) \\
\hline
BS8793 & 23$^h$ 07$^m$ 14.6$^s$& -50$^{\circ}$ 41$'$ 11.0$''$ &  5.83 &  0.8 & 8.5  \\
BS0897 & 02$^h$ 58$^m$  15.6$^s$ & -40$^{\circ}$ 18$'$ 17.0$''$ &  3.20 &  1.1 & 8.3  \\
BS1563 & 04$^h$ 50$^m$  55.1$^s$  & -53$^{\circ}$ 27$'$ 41.0$''$&  5.61&  0.8 & 12.3  \\
BS9002 &  23$^h$ 46$^m$  00.8$^s$ & -18$^{\circ}$ 40$'$ 41.0$''$ & 5.29  & 1.0  &  6.6 \\
BS1212 &  03$^h$ 54$^m$  17.4$^s$ &  -02$^{\circ}$ 57$'$ 17.0$''$ &   4.79 &   1.5 &  6.8 \\
BS2948&  07$^h$ 38$^m$  49.3$^s$ & -26$^{\circ}$ 48$'$ 07.0$''$ &  4.50 &   0.2 &  9.9 \\
BS0487 & 01$^h$ 39$^m$  47.7$^s$ &  -56$^{\circ}$ 11$'$ 41.0$''$ & 5.82 &  0.1 &  10.8 \\
\hline
\end{tabular}}
\caption{Binary stars observed during the PAR2007 site testing campaign (20 nights) extracted from the Bright Stars Catalogue. In the first column the name of the star, in the second and third columns the right ascension and the declination, in the fourth column the magnitude of the primary star, in the fifth column the difference in magnitude between the primary and secondary star of each binary, in the last column the angular separation.}
\label{tab_stars}
\end{table*}

The GS is a vertical profiler based on an optical remote sensing technique that is able to reconstruct the turbulence distribution developed on the whole atmosphere from the ground up to 20-25~km. The vertical resolution (\cite{vernin1983}) $\Delta$$H$ increases with the height $h$ and it depends on the binary separation, the wavelength and the geometrical optical set-up as:

\begin{equation}
\Delta H(h) = \frac{{0.78 \cdot \sqrt {\lambda | h - h_{gs}|} }} {\theta }  
\label{res}
\end{equation}

where $h$ is the height from the ground, $h$$_{gs}$ is the conjugated height under-ground (see Section \ref{instr}) and $\theta$ is the binary separation. Considering the typical values of the binaries used for the GS, we can say that the GS vertical resolution is typically of the order of 1~km. The GS is, at present, one of the most useful and reliable instrument for a deep and detailed characterization of the optical turbulence and its spatial distribution in the atmosphere because, with the \CN2 profiles and the wind speed\footnote{The wind speed can in principle be retrieved from the GS that has been applied with this goal in many studies (\cite{avila2001}, \cite{garcia2006}, \cite{avila2006}, \cite{egner2007}, \cite{garcia2009}, \cite{masciadri2010}). However, it is worth noting that it has been proved (\cite{masciadri2001}, \cite{hagelin2010}) that, wind speed profiles from meso-scale models can supply equivalent reliable estimates of this parameter with the advantage of a larger spatial and temporal cover. For a systematic turbulence monitoring this is a preferable solution.} profiles, it is possible to retrieve basically all the key parameters characterizing the optical turbulence for astronomical applications. The principle of the GS is solid, simple and many GS instruments have been developed in the last decade by different teams in the world. However, the GS is not a practical instrument for long terms and systematic monitoring of the turbulence and its distribution above an astronomical site because it requires a telescope with a pupil size of at least 1~m. For this application it is preferable to use automatic monitors such as the MASS, for example. The latter has a lower resolution ($\Delta$$h$ $\sim$ $h/2$) even if, recently, solutions to increase the resolution in the troposphere have been proposed  - \citep{kornilov2011}.

The necessity to recalibrate the GS measurements of the PAR2007 campaign appeared urgent after the recent studies (\cite{johnston2002} and \cite{avila2009}) that put in evidence an error in the procedure for retrieving the \CN2 profiles in the standard GS technique (in the post-processing phase) that could induce an overestimate of the turbulence. The error is perfectly described analytically therefore measurements can be corrected in a post-processing phase. In this paper we summarize the results obtained after the re-calibration of the data-set. A statistical analysis as well as an analysis on each single night and for each instant along all the nights is performed so to have an absolute unbiased reference for the GS measurements.

In Section \ref{instr} we describe the main principle of the GS technique, where the error of the GS technique has been identified and how to correct it. Effects of the errors induced on the vertical \CN2 profiles are discussed in Section \ref{cn2}. Effects of errors induced on the total seeing as well as the boundary and free atmosphere contributions are treated in Section \ref{see}. In Section \ref{iso} we treat the effects of errors induced on the isoplanatic angle. Besides, in Section \ref{res} a statistic of the median values of total seeing, boundary and free atmosphere seeing as well as isoplanatic angle obtained after re-calibration are provided. A brief discussion is presented in Section \ref{discus}. In Section \ref{concl} the conclusions of this study are presented.

\begin{figure*}
\begin{center}
\includegraphics[width=12cm]{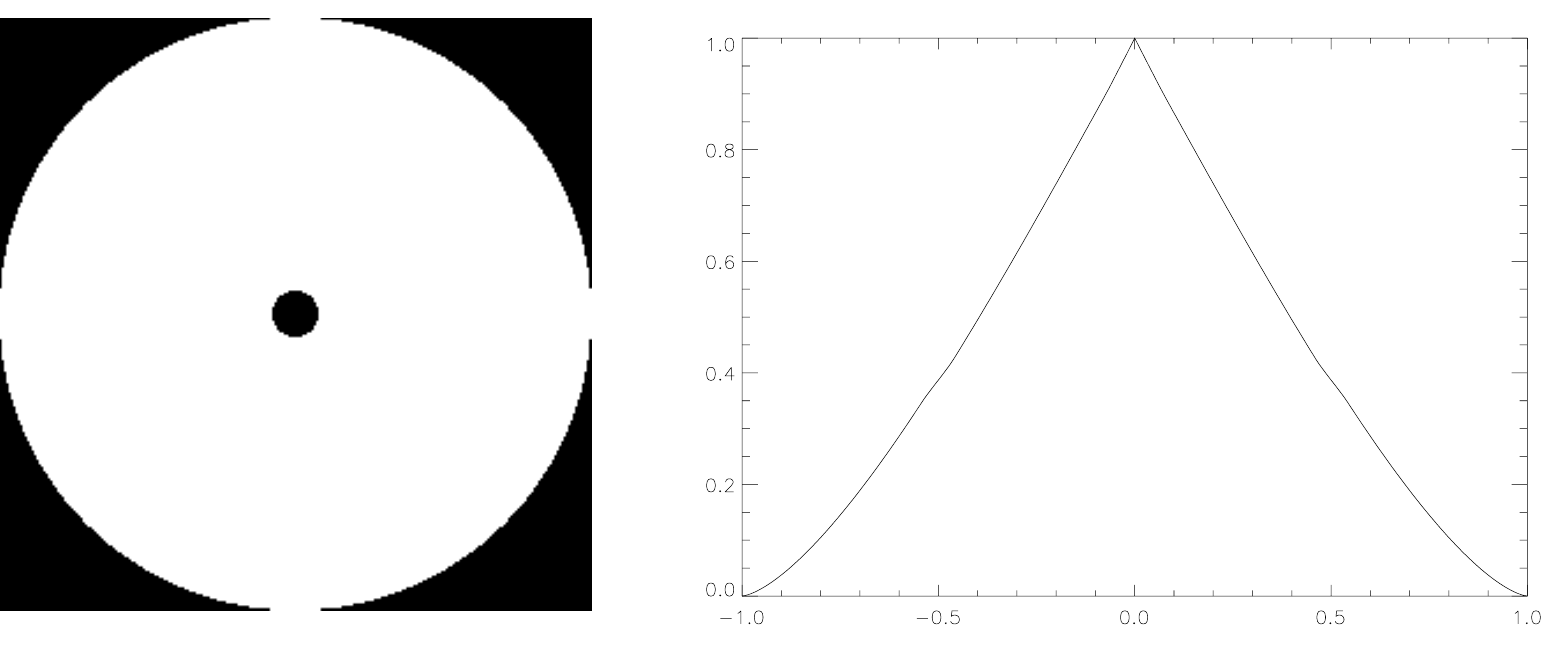}
\end{center}
\caption{Left: pupil of the Auxiliary Telescopes (ATs) of the VLT at Cerro Paranal: $D$=1.8 m, $e$=0.07. Right: autocorrelation of the telescope pupil shown on the left panel.}
\label{corr}
\end{figure*}

\section{The Instrument and the re-calibration}
\label{instr}

The SCIDAR technique (Scintillation Detection and Ranging) has been originally proposed by \cite{vernin1983} and \cite{rocca1974} to measure the vertical optical turbulence distribution (the refractive index structure constant \CN2 profiles) in the troposphere. The technique relies on the analysis of the scintillation images generated by binary stars on the pupil plane of a telescope. The standard SCIDAR technique (called Classic Scidar) is insensitive to the turbulence near the ground. This fact represented in the past an important limitation for monitoring turbulence for astronomical applications because it is known that most of the turbulence develops in the low part of the atmosphere. To overcome this limitation (\cite{fuchs1994}, \cite{fuchs1998}) proposed a generalized version of the SCIDAR (called Generalized SCIDAR - GS) in which the detector is virtually conjugated below the ground at a distance $h_{gs}$ permitting to extend the measurements range to the whole atmosphere (from the ground up to $\sim$ 20-25 km). GS measurements have been later done by several authors above different astronomical sites using instruments developed by several different teams (\cite{avila1997}, \cite{avila1998}, \cite{avila2004}, \cite{klueckers1998}, \cite{mckenna2003}, \cite{garcia2006}, \cite{egner2007}, \cite{egner_masciadri2007}, \cite{fuensalida2008}, \cite{masciadri2010}, \cite{garcia2011}, \cite{garcia2011b}, \cite{avila2011}). 

\begin{figure*}
\begin{center}
\includegraphics[width=14cm,angle=-90]{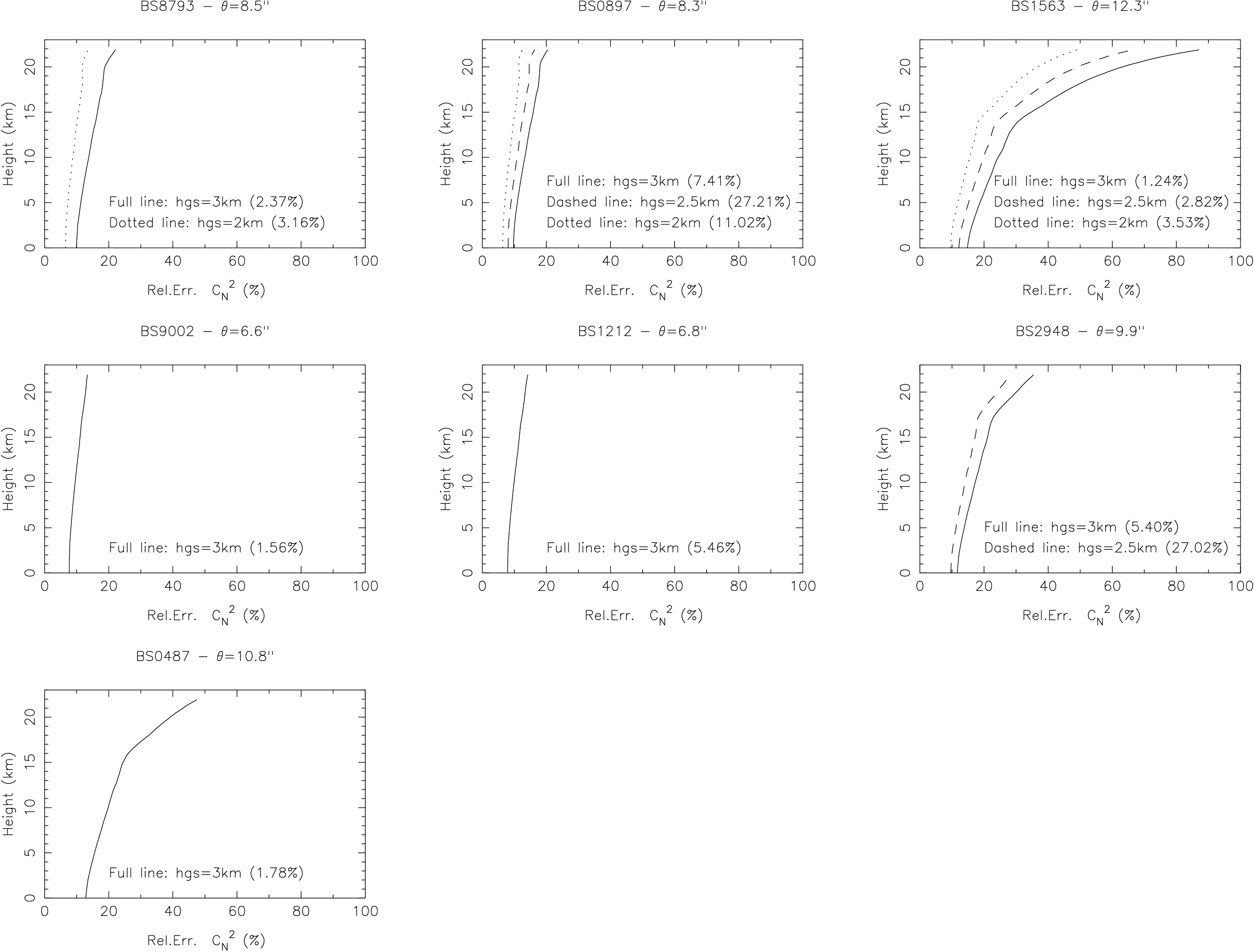}
\end{center}
\caption{Relative errors of the \CN2 vertical profiles ($\zeta(h)$) calculated for all the 20 nights monitored during the PAR2007 site testing campaign. The titles of the panels report the name of the observed binary stars with the correspondent angular separation. In each panel the relative errors with respect to the height $h$ associated to different values of the h$_{gs}$ (i.e. the height at which the CCD plane is conjugated underground) are shown. The numbers in parenthesis indicate the percentage of profiles with respect to the whole sample of measurements. }
\label{cn2_rel}
\end{figure*}

The GS PAR2007 measurements were taken from an Auxiliary Telescope (AT) of the VLTI with CUTE-Scidar III instrument, which has been developed by the Instituto de Astrof'sica de Canarias (IAC) team (V\'azquez-Rami\'o et al., 2008) to be used at the Paranal Observatory within the European project ÒELT design studyÓ. This instrument is an improved version of others previously developed by the same group for Roque de los Muchachos and Teide Observatories, which has been used since 2002 to extensively monitor the optical turbulence at both observatories (Fuensalida et al., 2004, \cite{garcia2011}, \cite{garcia2011b}). CUTE-Scidar III instrument provides \CN2 profiles in real time without dome and mirror seeing components. The binary stars specifications observed during the PAR2007 site testing campaign are listed in Table \ref{tab_stars}. Considering the combinations of values of $\theta$ and $h$$_{gs}$ used in the observations, we calculated that, using Eq.\ref{res}, the typical vertical resolution $\Delta$$H$ at $h$=0 is within 400-1000~m. The $\Delta$$H$(h) increases as Eq.\ref{res}.

The GS is based on the observation of binaries having an angular separation typically of $\sim$ 3-10 arcsec, magnitude $\le$ 5-6 mag and a $\Delta$$m$ $\sim$ 1 mag. When two plane wavefronts coming from a binary and propagating through the atmosphere meet a turbulent layer located at a height $h$ from the ground, they produce, on the detector plan, optically placed below the ground at a few kilometers ($h_{gs}$), two scintillation maps characterized by typical shadows appearing in couple at a distance $d$. Such a distance is geometrically related to the position of the turbulent layer as $d$ = $\theta$($h$+$h_{gs}$). The calculation of the auto-covariance (AC) of the scintillation map, normalized by the average of the intensity of the scintillation maps, produce the so called 'triplet'. The central peak is located in the centre of the AC frame; the lateral peaks are located at a symmetric distance $d$ from the centre. The amplitude of the lateral peaks is proportional to the strength of the turbulence of the layer located at the height $h$ weighted by the scintillation that such a layer produces on the detector. In a multi-layers atmosphere, different turbulent layers ($i$) produce triplets with lateral peaks located at different distances $d_{i}$ from the centre of the AC frame. The $\CN2$ profiles are obtained inverting the normalized function just described called Friedholm equation that describes several triplets. 

In 2002, Johnston et al. put in evidence an error in the calculation of the normalization of the AC frames and they studied the effects of this error on the \CN2 profiles for $h$ = 0. More recently, Avila \& Cuevas (2009), extended the analysis for heights $h$ $>$ 0. Results of this analysis say that, to obtain exact results for the \CN2 profiles at all heights $h$ from the ground, one has to multiply the $\CN2(h)$ retrieved from the GS by a factor (1/1+$\zeta(h)$) where $\zeta$ is the relative error between the exact and erroneous auto-covariance of the scintillation maps obtained with the GS and described in Eq.(\ref{eq1}):

\begin{equation}
\zeta (h) = \frac{{S(r - \theta h_{gs} )}}
{{aS(r) + b[S(r + \theta h_{gs} ) + S(r - \theta h_{gs} )]}} - 1
\label{eq1}
\end{equation}
where $S(r)$ is the auto-correlation of the pupil of the telescope, $a$ and $b$ are related to $\Delta$$m$ as Eq.(\ref{eq2}):
 
\begin{equation}
a = \frac{{1 + \alpha ^2 }}
{{(1 + \alpha )^2 }};\;b = \frac{\alpha }
{{(1 + \alpha )^2 }};\;\alpha  = 10^{ - 0.4\Delta m} 
\label{eq2}
\end{equation}

$\zeta$$(h)$ depends on a set of geometrical parameters related to the optical set up and the observed binaries, more precisely the pupil of the telescope $D$, the height $h_{gs}$ at which the detection plane is conjugated below the ground, the ratio between the stellar magnitudes of the binaries $\Delta$m, the angular separation of the binary $\theta$ and the ratio $e$ between the central obscuration $D^{*}$ and the telescope pupil ($e$=$D^{*}$/$D$). We obtain therefore $\zeta$=$\zeta$($\theta$,$h$$_{gs}$,$e$,$b$,$h$). Figure \ref{corr} shows the pupil size of the Auxiliary Telescopes of the Very Large Telescope where the GS has been run during the PAR2007 campaign and a section of the bi-dimensional autocorrelation of the pupil of the telescope $S$. We note that the value of $e$ for the AT at Paranal is smaller than in other telescopes used for GS measurements (\cite{masciadri2010},\cite{avila2011},\cite{garcia2011}). As it will be discussed in Sec.\ref{discus}, this is a favorable condition in the context of the error of the normalization of the scintillation maps of the GS.

\section{Results}
\label{res}
\subsection{Vertical turbulence distribution: \CN2}
\label{cn2}

\begin{figure}
\begin{center}
\includegraphics[width=5.5cm,angle=-90]{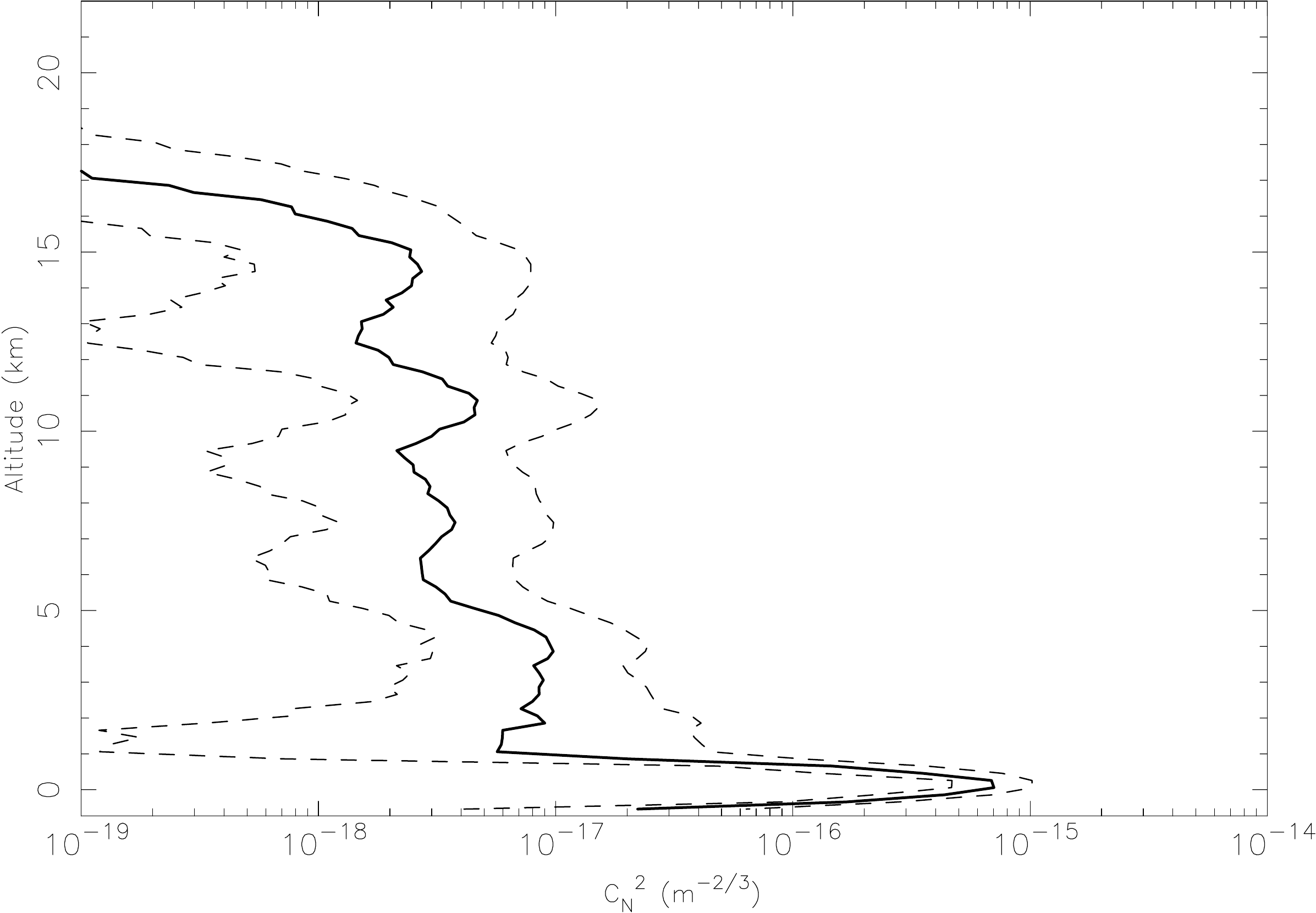}
\end{center}
\caption{Median vertical profile of the \CN2 calculated on the whole re-calibrated data-set of the 20 nights of the PAR2007 site testing campaign. Dashed lines represent the first and third quartiles.}
\label{cn2_median}
\end{figure}

\begin{figure*}
\begin{center}
\includegraphics[width=4cm,angle=-90]{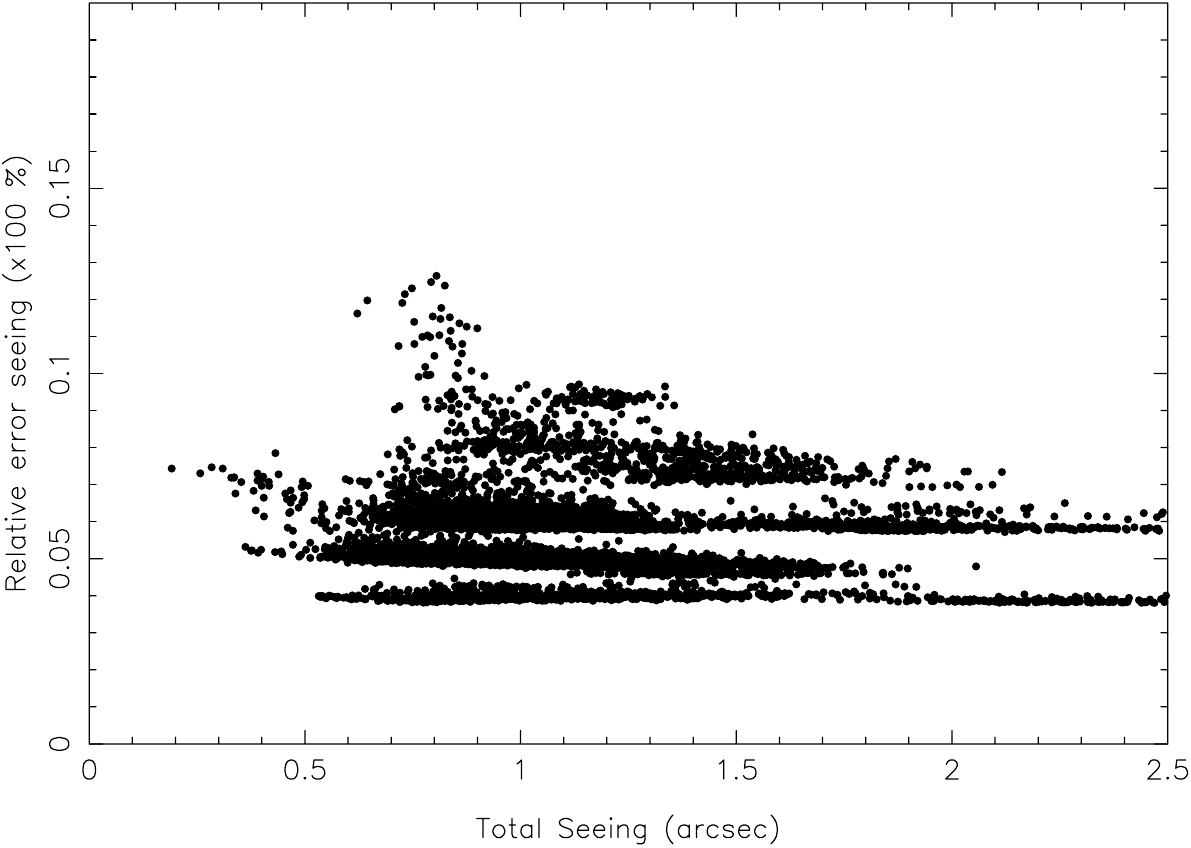}
\includegraphics[width=4cm,angle=-90]{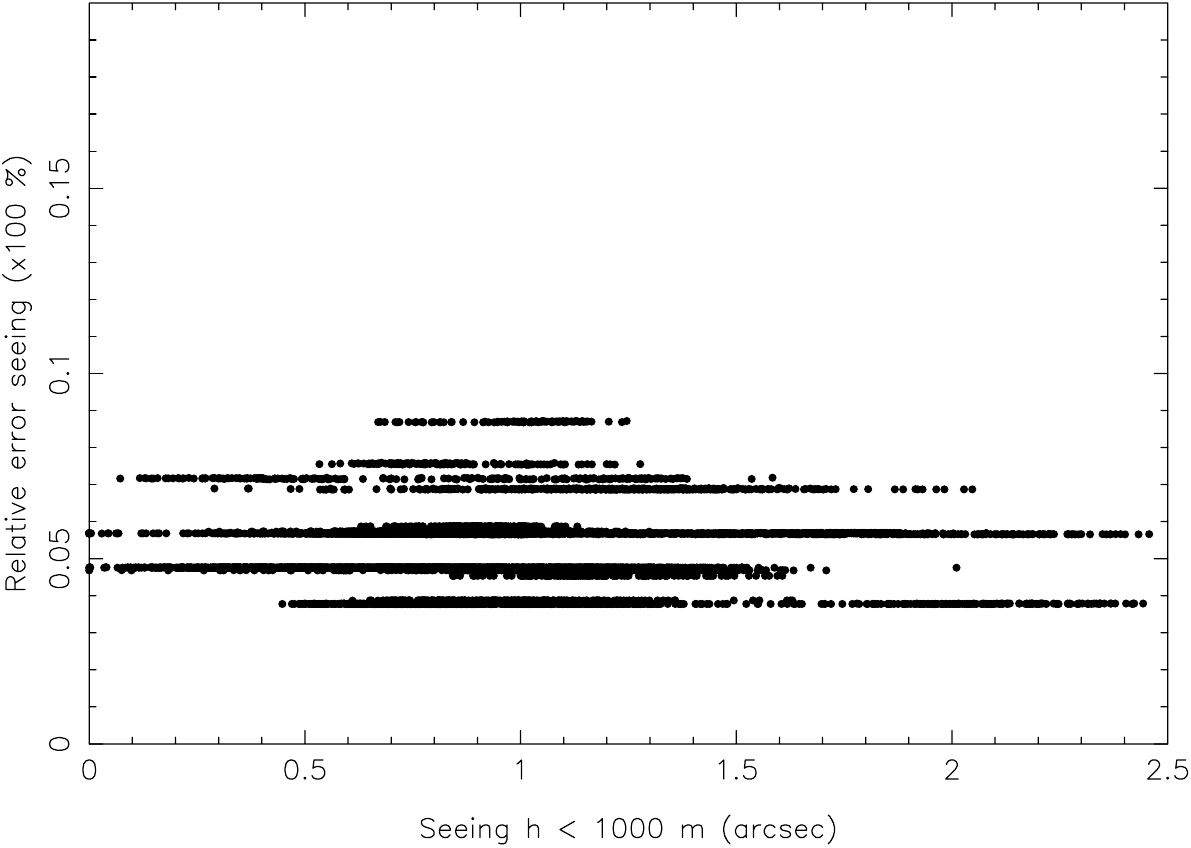}
\includegraphics[width=4cm,angle=-90]{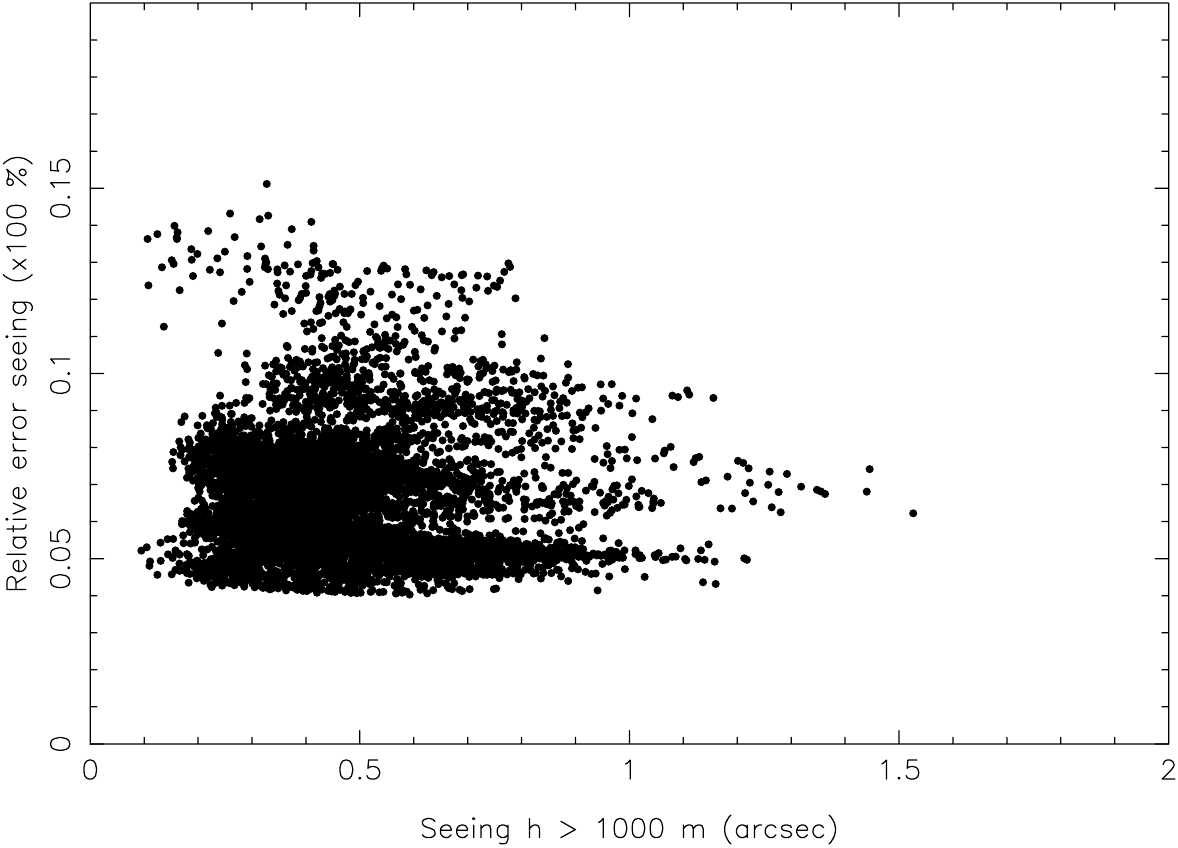}
\end{center}
\caption{Relative error versus total seeing (left), seeing below (centre) and above (right) 1000 m for all the nights of the PAR2007 site testing campaign. Y-axis values have to be multiplied by 100 and expressed in (\%). Ex: 0.1 means 10 \%.}
\label{see_rel}
\end{figure*}

\begin{figure*}
\begin{center}
\includegraphics[width=14cm]{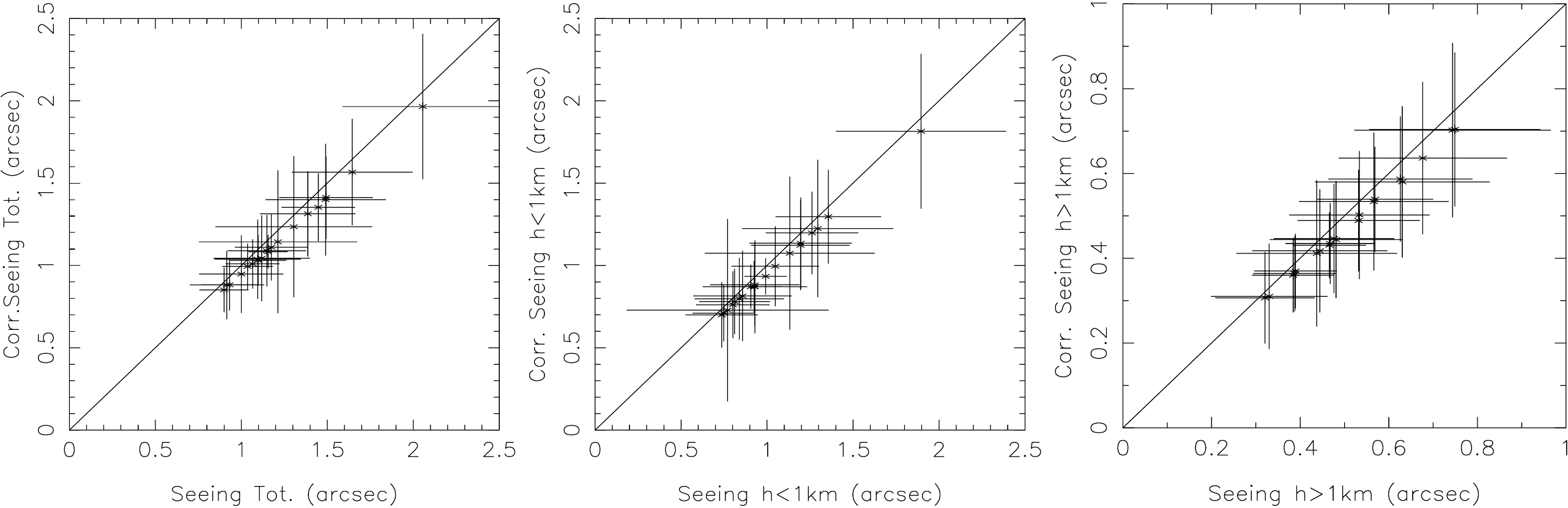}
\end{center}
\caption{Original (x-axis) versus re-calibrated (y-axis) seeing values for the 20 nights monitored during the site testing campaign of the PAR2007 site testing campaign. Each point is the average value of a single night. Left: total seeing. Centre: boundary layer seeing (h $<$ 1km). Right: free atmosphere seeing (h $>$ 1km). The error bars represent the standard deviation.}
\label{see_err}
\end{figure*}

Figure \ref{cn2_rel} reports the relative errors $\zeta(h)$ of the \CN2 profiles (before and after re-calibration) as a function of the height $h$ calculated for the 20 nights monitored during the PAR2007 site testing campaign. For each binary star (i.e. for a fixed angular separation $\theta$ and magnitude ratio $\Delta$$m$) the relative error is calculated for different values of the conjugated height under the ground ($h_{gs}$) (full, dashed and pointed lines). Panels in Fig.\ref{cn2_rel} include all the combinations of the set of parameters from which the error $\zeta$ depends on. Typically 3 or 4 binary stars are observed in each night and the position of the conjugated plane has been changed sometime during the nights. As a consequence, in each night we have the sequence of a few of the $\zeta(h)$ shown in Fig.\ref{cn2_rel}. We observe in Fig.\ref{cn2_rel} that $\zeta$ assumes a value of the order of 10$\%$ for $h$ =0 in almost all of the cases but it increases monotonically with the height reaching also values of the order of 40-60$\%$ in some cases above 15~km (BS1563). It is however worth noting that this happens for a small percentage of cases (see numbers in parenthesis in the panels of Fig.\ref{cn2_rel}). Indeed, if we look at the average \CN2 profiles obtained in each night before and after the re-calibration (Annex \ref{annex_avg_cn2}), we note that the vertical distribution (shape of the \CN2) remains substantially the same for each height and the difference between the original and the corrected \CN2 profiles appears small. This fact indicates that it is important to consider always both elements to appreciate the impact of the $\zeta(h)$ error on GS measurements.
Looking at Fig.\ref{annex_avg_cn2}, we note that the shape of the \CN2 profile is always characterized by a principal peak near the ground with most of the turbulent contribution. However, in the free atmosphere, we observe very different spatial distributions of the turbulence in each night and it is not necessarily visible the typical secondary peak at the jet-stream level that appears when we estimate the average or the median values of a rich statistical sample. This tells us that in the free atmosphere there is a large variability of the position of the layers in each night. As expected, in the re-calibrated \CN2 profiles, the turbulence is systematically weaker than in the original ones at all heights. In Annex \ref{annex_cn2-temp_evol} are reported the temporal evolution of the re-calibrated \CN2 profiles in each night of the PAR2007 campaign. Empty temporal windows correspond to an interruption of acquisition on the same binary star in some cases or to the change of the binary star in some other cases. Looking at Fig.\ref{annex_cn2-temp_evol} and Fig.\ref{temp_rel_see} (Section \ref{see}) it is possible to identify the instants in which the binaries have been changed and it is possible to discriminate between the two cases. In case the binary star is changed, this means that the parameters from which the relative error $\zeta(h)$ depends on are changed and therefore the value of the relative error changes sharply its value. We note that, most frequently, each night is characterized by a few layers (a sort of background structure) that are present during the whole night. Sometime it happens however, that a particular layer appears or disappears at a precise time during the night, for example on 10/11/2007 at 12~km at around 04:15 UT; on 17/12/2007 at 13~km at around 06:00 UT; on 19/12/2007 at 4~km at around 05:00 UT; on 20/12/2007 at 2~km at around 06:00 UT. Depending on the height and the intensity of each turbulent layers these events can induce more or less rapid and important changes of the integrated astro-climatic parameters. For example, the disappearing of a layer at 15~km can frequently produce, a weak decreasing of the total seeing (at this height the turbulent layers are in general much weaker than near the ground) but an important increase of the isoplanatic angle that is particularly sensitive to even small turbulence changes in the high atmosphere. 

Figure \ref{cn2_median} shows the median \CN2 profile obtained with the re-calibrated data-set associated to all the 20 nights of the PAR2007 campaign. In the free atmosphere we identify three major secondary peaks at 4~km, 7-8~km, 10-11~km and 14-15~km. 

\subsection{Integrated turbulence: seeing and J}
\label{see}

\begin{figure}
\includegraphics[width=6cm,angle=-90]{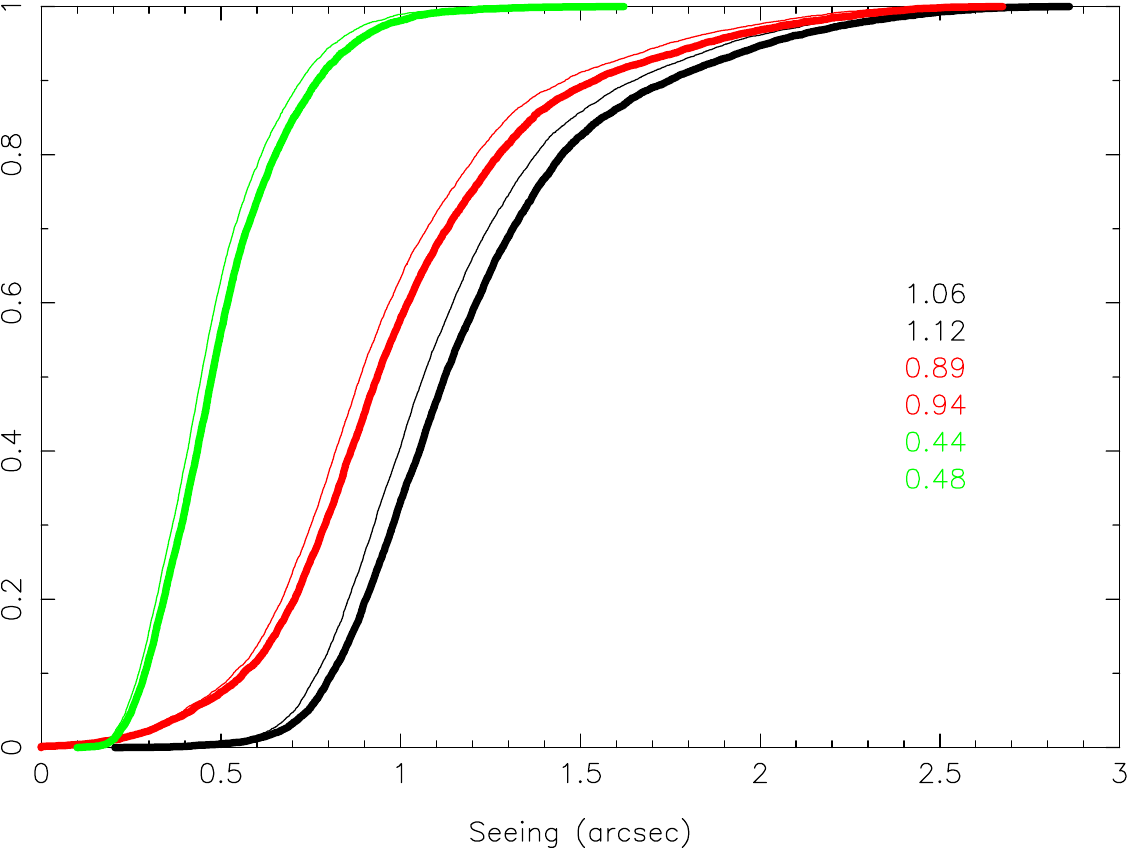}
\caption{Cumulative distribution of the total seeing (black line), seeing below (red line) and above (green line) 1000 m calculated before and after the re-calibration. The re-calibrated seeing (thin style line) is correctly always smaller than the original one (bold style line). The numbers in the panels are the median values calculated before (bold style line) and after (thin style line) the re-calibration.}
\label{cumdist_all}
\end{figure}

In Annex \ref{temp_rel_see} are reported the temporal evolution of the relative errors of the total seeing $\varepsilon_{TOT}$, the boundary $\varepsilon_{BL}$ and the free atmosphere seeing $\varepsilon_{FA}$ (before and after re-calibration) for all the 20 nights of the PAR2007 campaign. The relative error for the total seeing is defined as:

\begin{equation}
r_{TOT}  = \frac{{\varepsilon _{TOT}^{}  - \varepsilon _{TOT}^* }}
{{\varepsilon _{TOT}^* }}
\label{eq4}
\end{equation}
\noindent
where $\varepsilon _{TOT}^*$ and $\varepsilon _{TOT}$ are:

\begin{equation}
\varepsilon _{TOT}^*  = A(\lambda ) \cdot \left( {\int\limits_0^{20km} {C_N^2 (h)\frac{1}
{{1 + \zeta (h)}}dh} } \right)^{3/5} 
\end{equation}

\begin{equation}
\varepsilon _{TOT}^{}  = A(\lambda ) \cdot \left( {\int\limits_0^{20km} {C_N^2 (h)dh} } \right)^{3/5} 
\end{equation}
and $A$($\lambda$) is equal to 19.96$\times$10$^{6}$ for $\lambda$=0.5 $\mu$m.
Similar equations have been built for the relative error for $\varepsilon_{BL}$ and $\varepsilon_{FA}$ replacing the extremes of the integrals with the ranges: [0-1000m] and [1000m,20km]. Looking at Fig.\ref{temp_rel_see}, we see that, in all cases, the relative error of the boundary layer is relatively modest, most of time of the order of 5-9$\%$. The relative error of the free atmosphere seeing is slightly greater but always smaller than 15$\%$. It is greater than 10$\%$ only in 5 occasions (16/11/2007, 17/11/2007, 14/12/2007, 16/12/2007) and for a partial part of the night. We note that, for the boundary layer contribution, the relative error is almost constant and it changes its values basically when the binary star is changed. On the other side, the relative errors of the free atmosphere fluctuate in time (sometimes in a not negligible way but always more than in the boundary layer where the relative error remains basically constant in each segment associated to each observed binary). Why ? It is true that the absolute error (Eq.\ref{eq1}) increases with the height but this means that the relative errors at each instant and the successive one are both larger in the high part of the atmosphere than in the low part of the atmosphere. Therefore the fact that we note a systematic and more evident variation of the relative error in successive instants in the free atmosphere and not in the boundary layer can not be related to this reason. This is highly probably due to the fact that the \CN2 in the free atmosphere is, in absolute terms, a couple of order of magnitude smaller than the \CN2 near the ground.  As a consequence, when we calculate the relative error in the free atmosphere, we find that small variations in the \CN2 are associated to larger difference between the relative error at an instant $t$$_{0}$ and the successive one. In other words, the variation of the numerator of Eq.\ref{eq4} at two successive instants is much more smaller than the denominator in the case of the boundary layer than in the case of the free atmosphere. Figure \ref{see_rel} shows the relative errors for the total seeing, the seeing in the boundary layer and in the free atmosphere as a function of the corresponding values of seeing for all the nights of the PAR2007 site testing campaign. It is evident that no correlation exist between the relative error and the corresponding value of the seeing. This means that it is not possible to retrieve any relation between the value of the relative error of the seeing and the value of the seeing. However, it is possible to note that the fluctuations of the relative errors in the free atmosphere are larger than in the boundary layer and the distribution of the relative errors of the seeing assumes a sort of 'cloud shape'. Figure \ref{see_err} shows the re-calibrated averaged values of the seeing calculated in each nights versus the original averaged values. As expected, a systematic decreasing of the estimated turbulence in the re-calibrated data is evident. At the same time, we observe that the difference on the mean values (before and after the re-calibration) is very small if compared to the standard deviation of each estimate. The greater is the seeing, the greater is the absolute difference between the average seeing before and after re-calibration. Figure \ref{cumdist_all} shows the cumulative distribution of the total seeing, seeing in the boundary layer and in the free atmosphere before and after the re-calibration. The corresponding median values reported in Fig.\ref{cumdist_all} tell us that the absolute errors for the median values of the seeing are: 
$\Delta$$\varepsilon_{BL}$=0.05 arcsec, $\Delta$$\varepsilon_{FA}$=0.04 arcsec, $\Delta$$\varepsilon_{TOT}$=0.06 arcsec. In Annex \ref{see_temp_evol} is shown the temporal evolution of the seeing (total, boundary layer and free atmosphere for each night. 

\begin{figure*}
\begin{center}
\includegraphics[width=4cm,angle=-90]{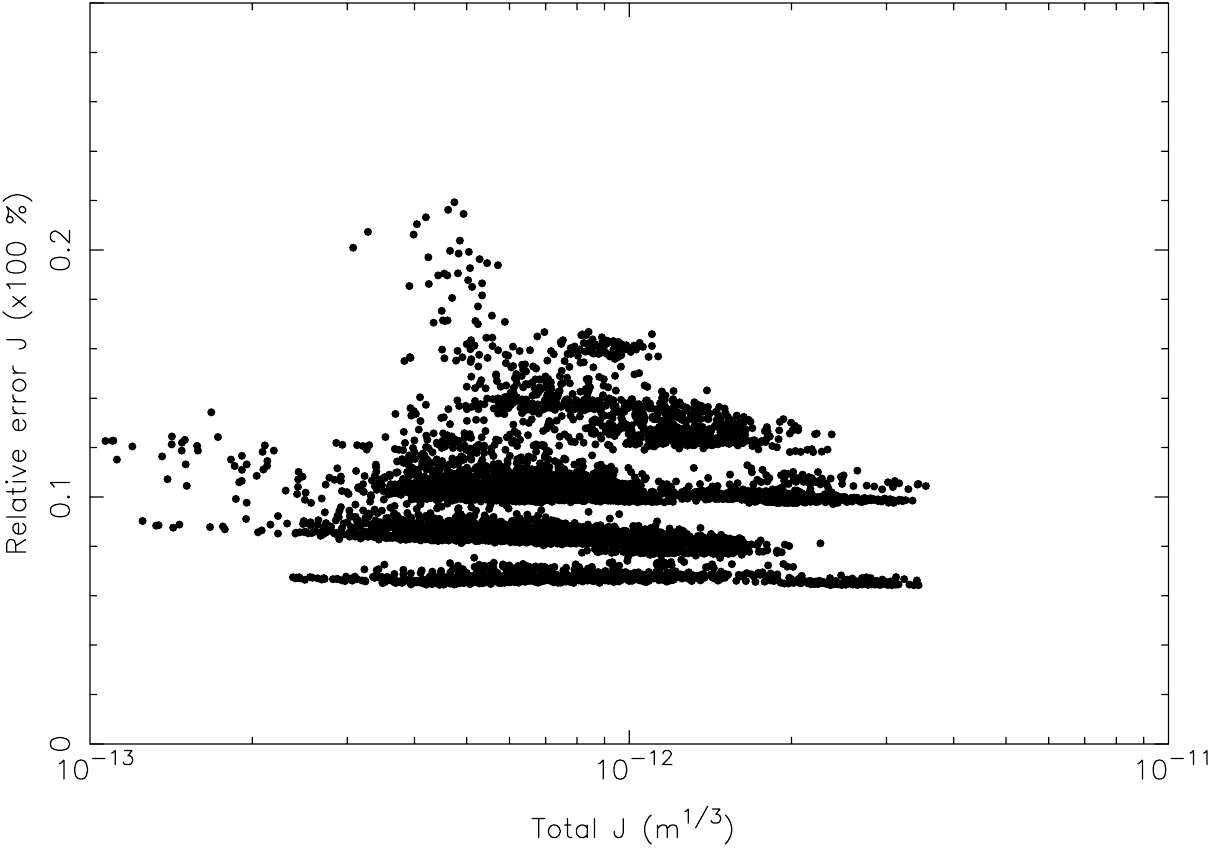}
\includegraphics[width=4cm,angle=-90]{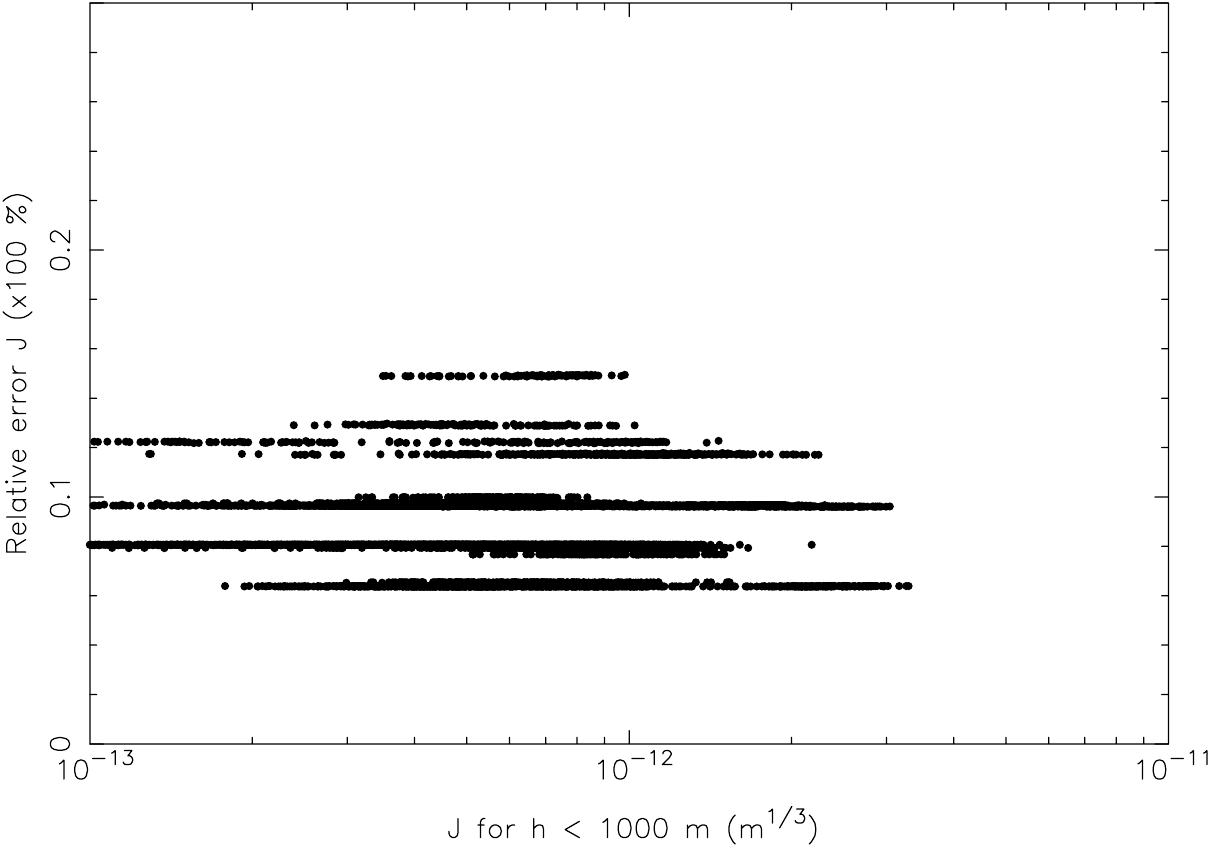}
\includegraphics[width=4cm,angle=-90]{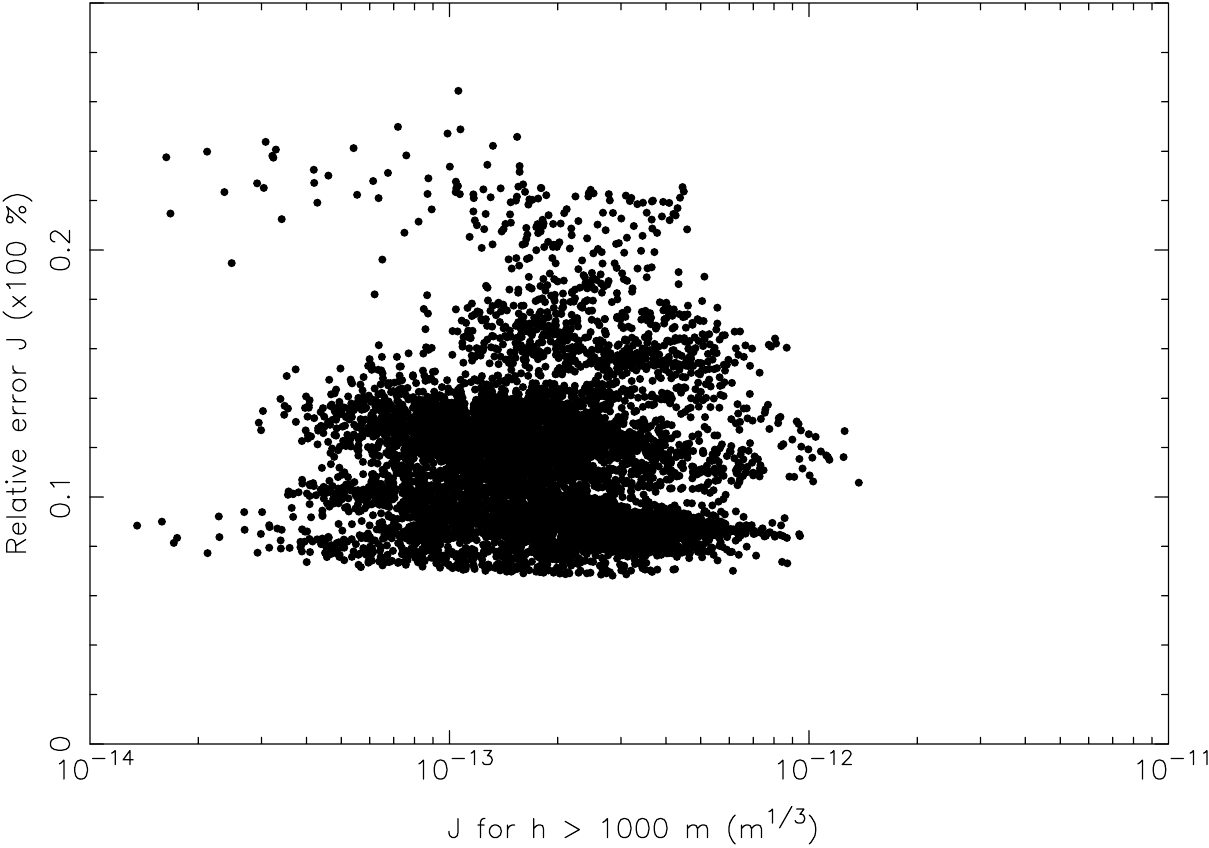}
\end{center}
\caption{Relative error versus total J (left), J below (centre) and above (right) 1000 m for all the nights of the PAR2007 site testing campaign. Y-axis values have to be multiplied by 100 and expressed in ($\%$). Ex: 0.1 means 10 $\%$.}
\label{j_rel}
\end{figure*}

It is worth noting (Fig.\ref{j_rel}) that the relative error of the total energy $J$ (measured in m$^{1/3}$) defined as:
\begin{equation}
J  =  {\int\limits_0^{20km} {C_N^2 (h)dh} } 
\end{equation}

is higher than that of the seeing and it can reach 20$\%$ in the free atmosphere, as well as in the total atmosphere. We distinguished between $J$ and seeing because, in some applications, it can be preferable to treat the total energy instead of the seeing. $J$ is indeed linear with respect to addition while seeing is not.

\begin{figure*}
\begin{center}
\includegraphics[width=5cm,angle=-90]{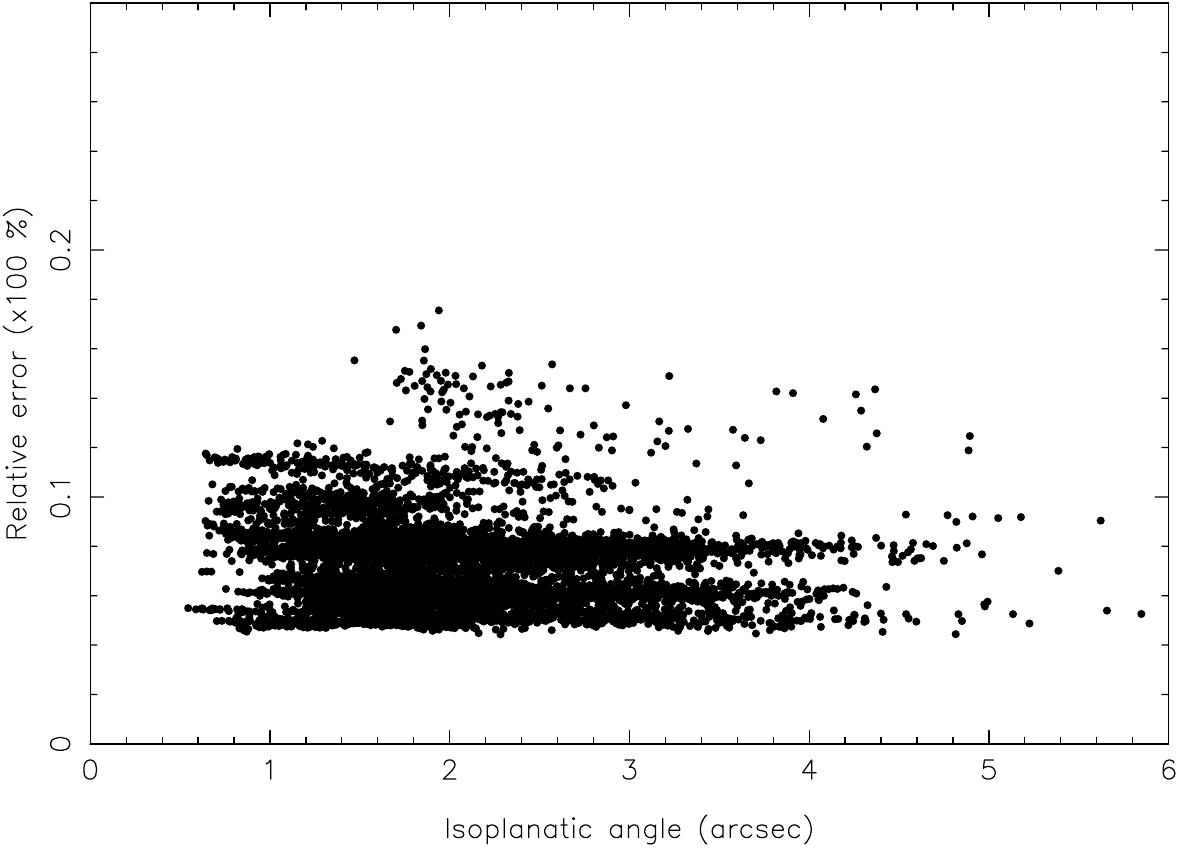}
\includegraphics[width=5cm,angle=-90]{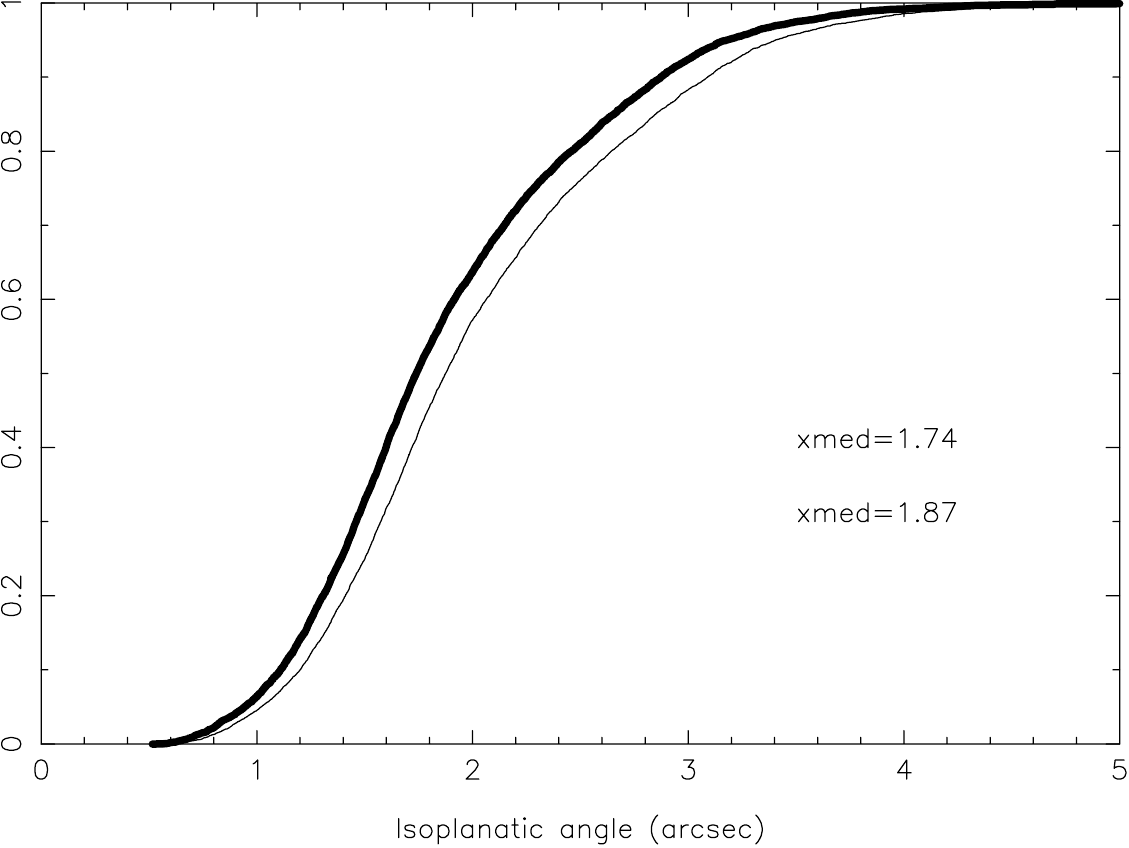}
\end{center}
\caption{Relative error versus isoplanatic angle (left) and cumulative distribution of the isoplanatic angle ($\theta_{0}$) before (bold style line) and after (thin style line) the re-calibration of the data (right).}
\label{iso1}
\end{figure*}

\subsection{Isoplanatic angle}
\label{iso}

In Annex \ref{annex_iso}  is reported the temporal evolution of the relative error of the iso-planatic angle ($\theta_{0}$) for all the 20 nights of the PAR2007 campaign. $\theta_{0}$ is defined as:

\begin{equation}
\theta _0  = 0.057 \cdot \lambda ^{6/5} \left( {\int\limits_0^{h_{up} } {h^{5/3} C_N^2 (h)dh} } \right)^{ - 3/5} 
\end{equation}
\noindent
where $h$$_{up}$ corresponds to the top of the atmosphere (more frequently to the end of vertical profile). We observe that the features of the temporal evolution of the $\theta_{0}$ relative error are similar to those of the free atmosphere seeing. This is not surprising because the isoplanatic angle is particularly  sensitive to the turbulence in the high part of the atmosphere due to the presence of the factor $h$$^{5/3}$ in the integral. The $\theta_{0}$ relative error is never larger than 18$\%$ with the exception of a few minutes on 17/11/2007 in which the error was of the order of 18$\%$. Figure \ref{iso1}-left shows the $\theta_{0}$ relative error versus the corresponding value of  $\theta_{0}$ related to all the 20 nights of the PAR2007 site testing campaign. Also in this case, as well as for the seeing, no correlation between the two quantities is observed. Figure \ref{iso1}-right shows the cumulative distribution of the iso-planatic angle before and after re-calibration. As expected, the median value of the distribution increases after the re-calibration (from 1.74 arcsec to 1.87 arcsec) because the total turbulence decreases. We highlight that $\theta_{0}$ in Fig.\ref{iso1} is calculated using $h$$_{up}$=20 km (i.e. 20~km from the ground). Previous studies in the literature (\cite{avila2004}, \cite{masciadri2004}, \cite{egner2007}, \cite{masciadri2010}, \cite{hagelin2011}, \cite{garcia2011}, \cite{garcia2011b}) performed the calculation of $\theta_{0}$ with a value of $h$$_{up}$ between 20 and 22~km from the ground. To estimate the effect of the value of $h$$_{up}$ on $\theta_{0}$ and to evaluate potential differences due this element, we calculated the integral using  $h$$_{up}$=22~km. We find, respectively, 1.73 arcsec and 1.87 arcsec. We conclude, therefore, that the median value of $\theta_{0}$ remains the same (1.87 arcsec) within one hundredth of arcsec in these data-set. In Annex \ref{iso_temp_evol} is shown the temporal evolution of the isoplanatic angle for each night.

\section{Discussion}
\label{discus}

From a qualitative point of view we can say that, as shown in Avila \& Cuevas (2009), the greater is the distance of the conjugated plane under ground ($h_{gs}$), the greater is $\zeta(h)$; the greater is the binary separation $\theta$, the greater is $\zeta(h)$; the greater is $\Delta$$m$, the greater is $\zeta(h)$; the smaller is the pupil size $D$, the greater is $\zeta(h)$; the greater is the ratio between the central obscuration and the pupil size $e$=$D^{*}$/$D$, the greater is $\zeta(h)$ near the ground. What about the quantitative effects ? We put in the context our results with respect to others recently published.

\cite{masciadri2010}, treating a sample of measurements related to 43 nights done above Mt. Graham (VATT telescope - $D$=1.83~m),  proved that GS measurements obtained with a pupil size $D$ $\ge$ 1.83~m and a binary separation $\theta$ $\le$ 8 arcsec are affected by this error for less than a few hundredths of an arcsec (0.04 arcsec corresponding to a relative error of $\sim$ 5$\%$ for the total seeing) with an average \CN2 profile reproducing the same spatial distribution of the un-corrected profile. 

Similar results have been obtained by Avila et al. (2011) treating measurements done above the San Pedro M\'artir Observatory ($D$ =1.5~m and $D$=2.1~m) in which it has been estimated a typical relative error of 4.2$\%$ (an absolute error of $\sim$ 0.04 arcsec on the total seeing). Modest errors therefore even if, in that case, a binary with $\theta$ = 14.4 arcsec ($\zeta$ Uma) was selected among the targets. The modest relative error is due to the fact that the percentage of time of observation of $\zeta$ Uma  was relatively low (14.4 $\%$).  

Results of this paper for measurements done at Cerro Paranal with a pupil size $D$=1.8 m are coherent and similar to the previous ones. A relative error for the total seeing of 5.6$\%$ (absolute error equal to 0.06 arcsec) and a relative error of the isoplanatic angle of 6.9$\%$ (absolute error equal to 0.13 arcsec) are calculated. In just one case a binary separation larger than 10 arcsec has been used (BS1563 - $\theta$=12.3 arcsec - percentage of observation is 7$\%$).

Some larger errors have been observed in the study performed at Roque de los Muchachos Observatories on Canary Islands (\cite{garcia2011b}) because a set of different geometric parameters (wider binary separation and deeper conjugation heights of the detector plane) have been used in combination with a smaller telescope pupil size.  In the study of Roque de los Muchachos ($D$=1~m), five binaries with angular separation wider than 10 arcsec (for a percentage of time equal to 3.39$\%$) have been selected. The absolute error of the median values of the total seeing is of the order of  0.12-0.15 arcsec and that of $\theta$ of the order of 0.34-0.92 arcsec (for observations at high and low vertical resolution). We conclude that, considering the relative small percentage of observations with wide binaries, the element that mostly affected the overestimate of the turbulence in the study done at Roque de los Muchachos was the small size of the pupil of the telescope ($D$=1~m) and a deep conjugation heights of the detector plane. However, looking at Fig.5 (\cite{avila2009}), we observe that, from a quantitive point of view, the impact of the telescope pupil size is much more important than the conjugation heights of the detector plane. The size of the telescope seems therefore to be the most critical parameter in the error of the normalization of the average of the scintillation map in the GS measurements published so far. To contain the relative error it is suggested  to keep the value of the conjugated height under ground $h_{gs}$ at 2 or 3 kilometers maximum  and to use a telescope with a pupil size of at least 1.5 m. The binary separation affected less the statistical results published so far because the number of measurements obtained with wide binaries was relatively small. 

\section{Conclusion}
\label{concl}

In this paper a re-calibration of the whole data-set of measurements performed with the Generalized Scidar (CUTE-Scidar) at Cerro Paranal on November/December 2007 (PAR2007 site testing campaign) for 20 nights is performed. The statistical analysis of the spatial and temporal distribution of the optical turbulence (\CN2 profiles) indicates relative errors that can achieve order of 40-60$\%$ (Fig.\ref{cn2_rel}) in some part of the atmosphere and in some narrow temporal windows. However, when we calculate the relative errors of the integrated counterparts, the relative error reach a maximum value of 13$\%$ for the correspondent total seeing $\varepsilon_{TOT}$ and 15$\%$ for the free atmosphere. The absolute errors are much smaller, particularly if one considers the \CN2 average on a whole night. The shape of the average of the \CN2 profiles before and after re-calibration is basically the same. For what concerns the integrated astroclimatic parameters, from a statistical point of view, we find a relative error of the median value of the total seeing equal to 5.6$\%$, for the seeing in the boundary layer equal to 5.6$\%$ and for the seeing in the free atmosphere seeing equal to 9$\%$. These values correspond to absolute errors respectively equal to $0.06$ arcsec, $0.05$ arcsec and $0.04$ arcsec. The absolute errors provided by the boundary layer and the free atmosphere contributions are statistically comparable even if the relative error increases with the height. The relative error of the turbulent energy $J$ is slightly larger. It can reach a maximum of 21$\%$ in the total atmosphere and 23$\%$ in the free atmosphere. The relative errors for the isoplanatic angle are never larger than 18$\%$. The absolute error of the median value of $\theta_{0}$ is $0.13$ arcsec. The re-calibration of the GS data-set performed in this paper suggests a revision of results obtained in \citep{dali2010} on the cross-comparison between optical turbulence measurements obtained with different instruments during the PAR2007 campaign. A few of us are, at present, deeply involved in a cross-checking of such a measurements with predictions performed with an atmospherical non-hydrostatic numerical model and a forthcoming paper will be dedicated to this topic. 

Results of this paper, joint with results found on previous papers on the same topic, indicate that, if a GS is used with a pupil size $D$ $\ge$ 1.8~m and an angular separation smaller than 10 arcsec (preferably with a value of the conjugated plane not deeper than $\sim$ 3~km underground), the absolute errors can be considered negligible (of the order of a few hundredth of arcseconds) from a statistical point of view. 

If $D$ $<$ 1.8~m and/or the angular separation is greater than 10 arcsec, the re-calibration is highly suggested, in particular if one is interested on some specific temporal windows. We remember that it has been proved \citep{masciadri2010} that the GS-HVR technique \cite{egner_masciadri2007} to be applied in first kilometer from the ground is not sensible to the error of the normalization of the autocorrelation. This solution can therefore be selected to investigate the turbulence at high vertical resolution near the ground (first kilometer) without necessity to re-calibrate the measurements. 

\section*{Acknowledgments}
This study is funded by the ESO contract number E-SOW-ESO-245-0933 (MOSE Project - PI: E. Masciadri). Measurements of the CUTE-Scidar - PAR2007 site testing campaign have been taken by the IAC Team (M.N. Cagigal, J.M. Delgado, J.J. Fuensalida, E. Hern\'andez, M. Reyes, H. V\'azquez Rami\'o) leaded by J.J. Fuensalida in collaboration with the ESO Astroclimatology Group (G. Lombardi, J. Navarrete, M. Sarazin) leaded by M. Sarazin, in the context of the project ELT-TRE-UNI-12300-0001 (ELT Design Study). CUTE-Scidar III instrument was partially funded by the Instituto de Astrof'sica de Canarias and by the Spanish Ministerio de Educaci—n y Ciencia (AYA2009-12903). The authors acknowledge R. Avila for the interesting discussions on the FFT.

%
%

\clearpage
\appendix

\section{Average \CN2 profiles}
\label{annex_avg_cn2}

\begin{figure*}
\begin{center}
\includegraphics[width=15cm]{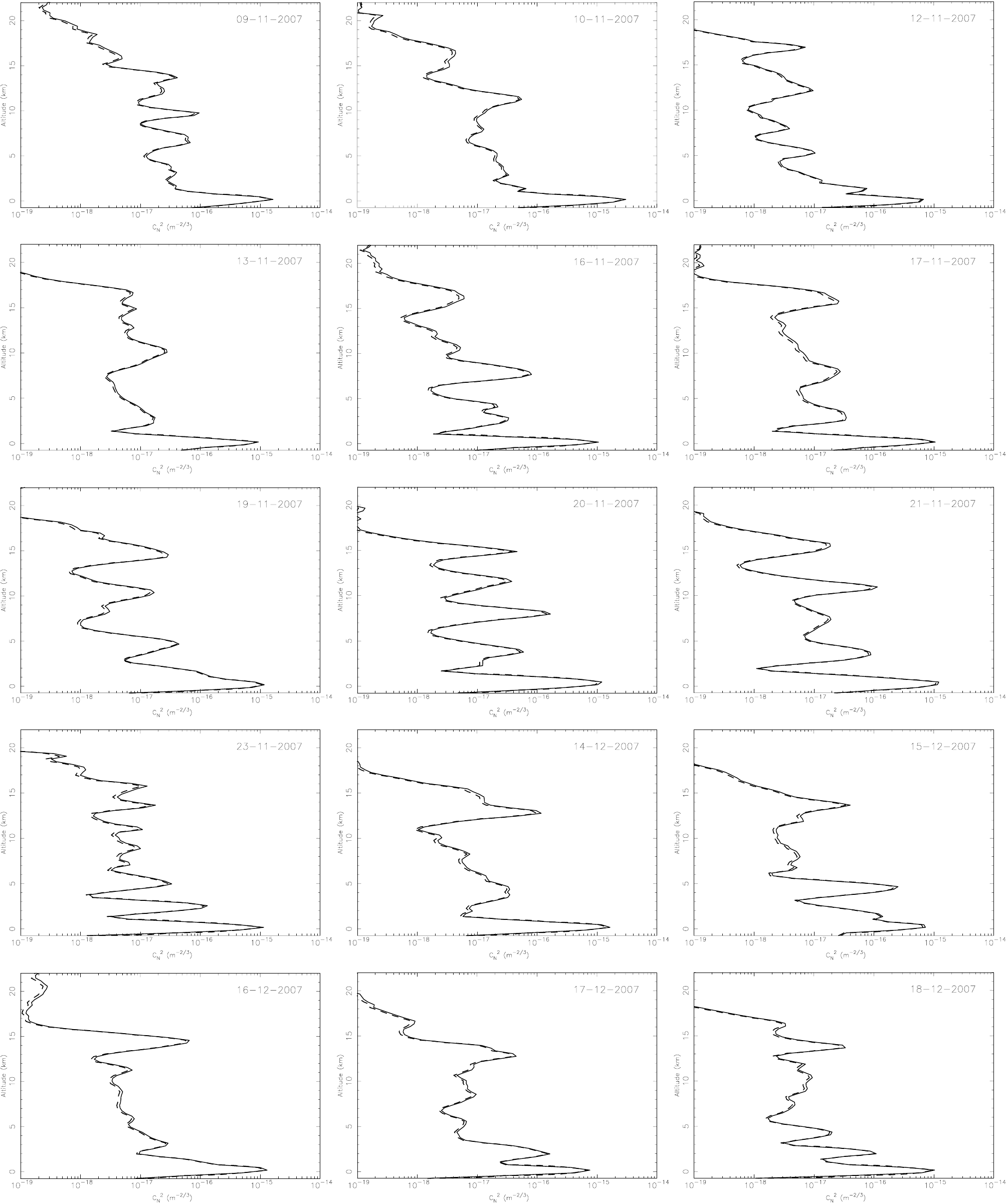}
\end{center}
\caption{Average \CN2 vertical profiles for each night of the PAR2007 site testing campaign obtained with the original data-set (full line) and the re-calibrated data-set (dashed line). Ex: 9/11/2007 corresponds to 10/11/2007 in UT. Error values are expressed in percentage. 
}
\label{profili_cn2_LT_1}
\end{figure*}

\begin{figure*}
\begin{center}
\includegraphics[width=15cm]{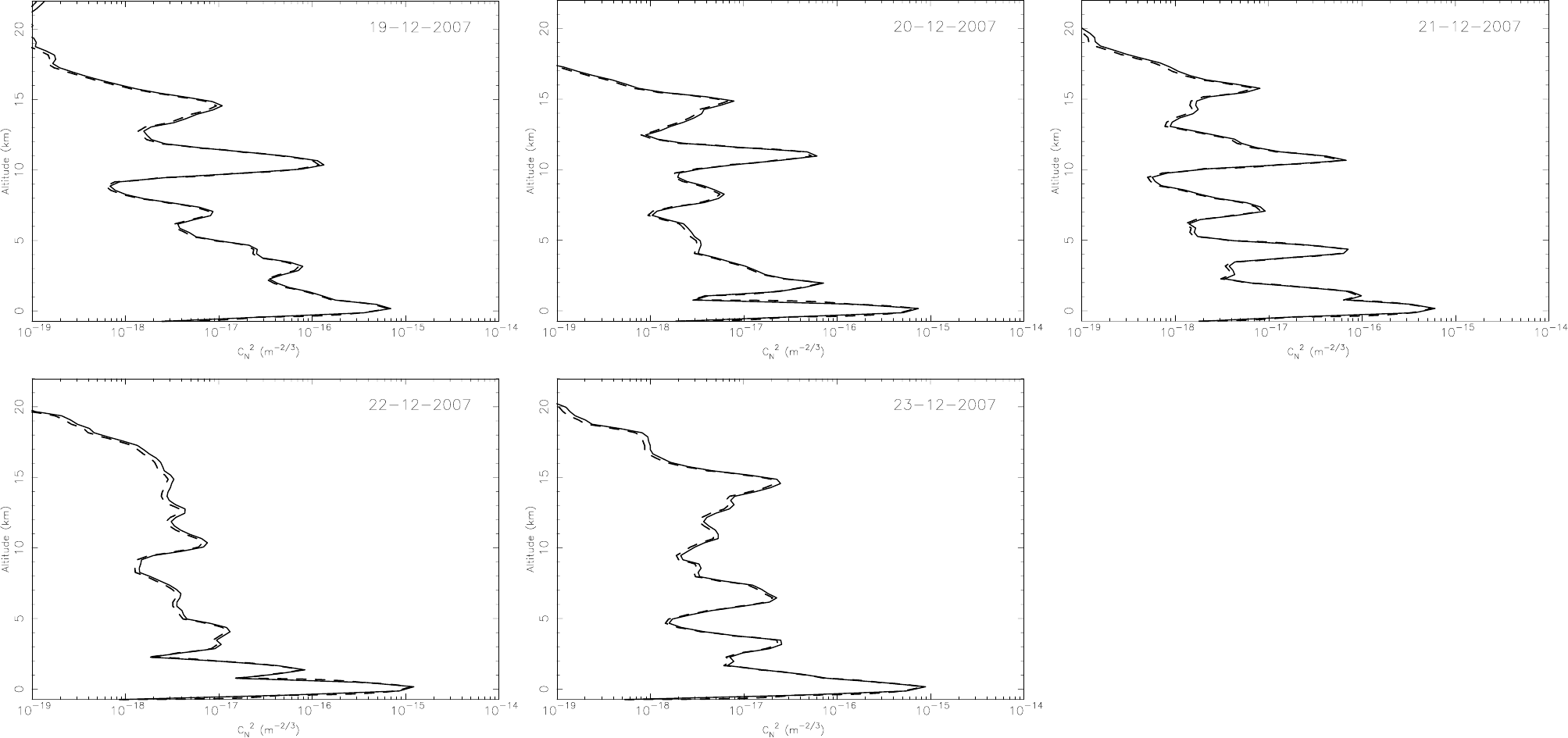}
\end{center}
\caption{It follows Fig. \ref{profili_cn2_LT_1}}
\label{profili_cn2_LT_2}
\end{figure*}

Fig.\ref{profili_cn2_LT_1} and Fig.\ref{profili_cn2_LT_2} show the average \CN2 profiles for each night of the PAR2007 campaign before and after the data-set re-calibration. The dates are in local time (LT). 

%
%

\section{\CN2 profiles temporal evolution}
\label{annex_cn2-temp_evol}

\begin{figure*}
\begin{center}
\includegraphics[width=15cm]{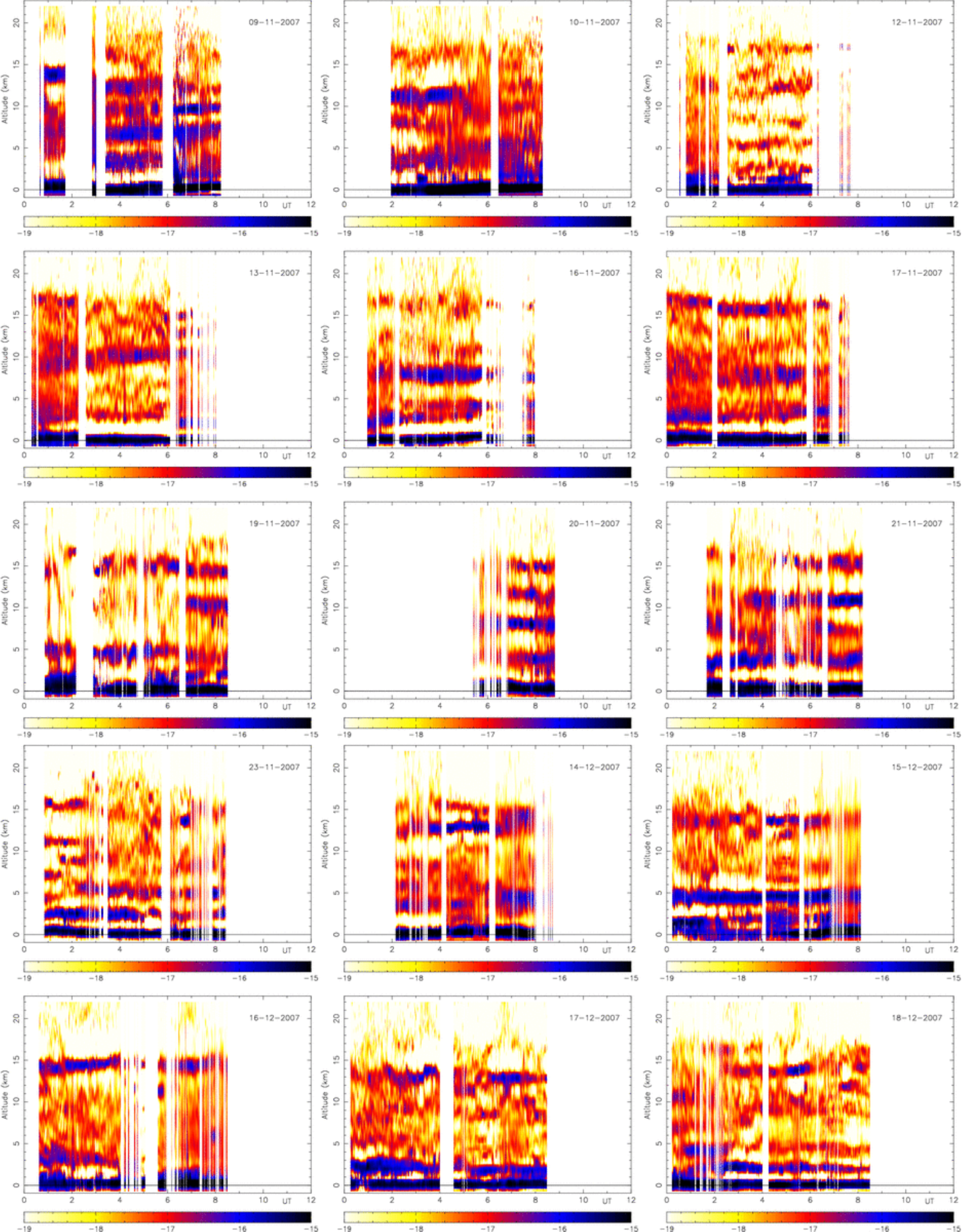}
\end{center}
\caption{Temporal evolutions of re-calibrated \CN2 vertical profiles for all the 20 nights monitored during the PAR2007 site testing campaign (units are $m^{-2/3}$ and in log scale). The dates are in local time (LT). Ex: 9/11/2007 in LT corresponds to 10/11/2007 in UT.}
\label{evol_temp_cn2_p1}
\end{figure*}

\begin{figure*}
\begin{center}
\includegraphics[width=15cm]{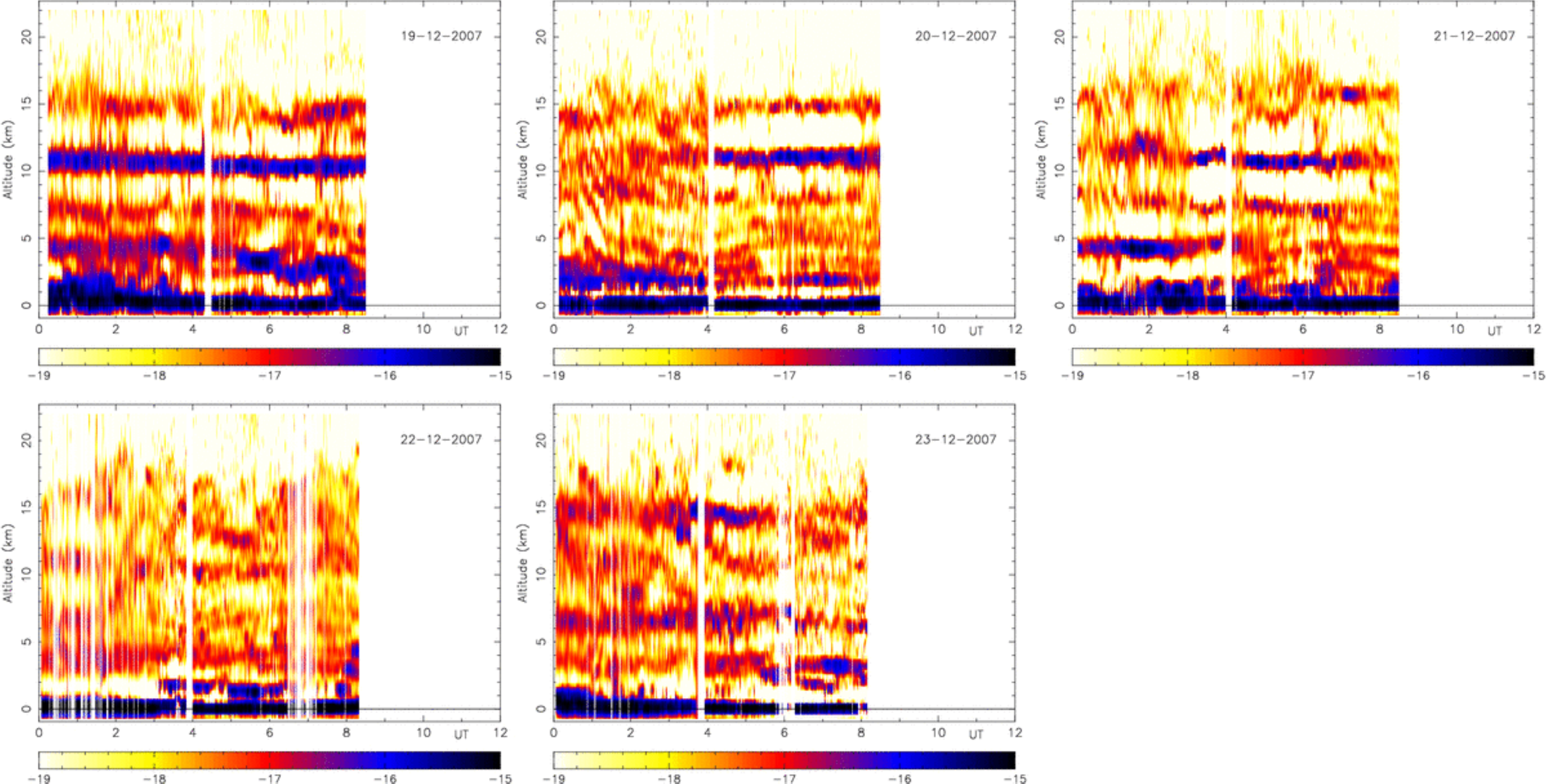}
\end{center}
\caption{It follows Fig. \ref{evol_temp_cn2_p1}.}
\label{evol_temp_cn2_p2}
\end{figure*}

Fig.\ref{evol_temp_cn2_p1} and Fig.\ref{evol_temp_cn2_p2} show the temporal evolution of the \CN2 profile during the night for all the 20 nights of the PAR2007 site testing campaign.

%
%

\section{Temporal evolution of the relative error of the seeing for the individual nights}
\label{temp_rel_see}

\begin{figure*}
\begin{center}
\includegraphics[width=15cm]{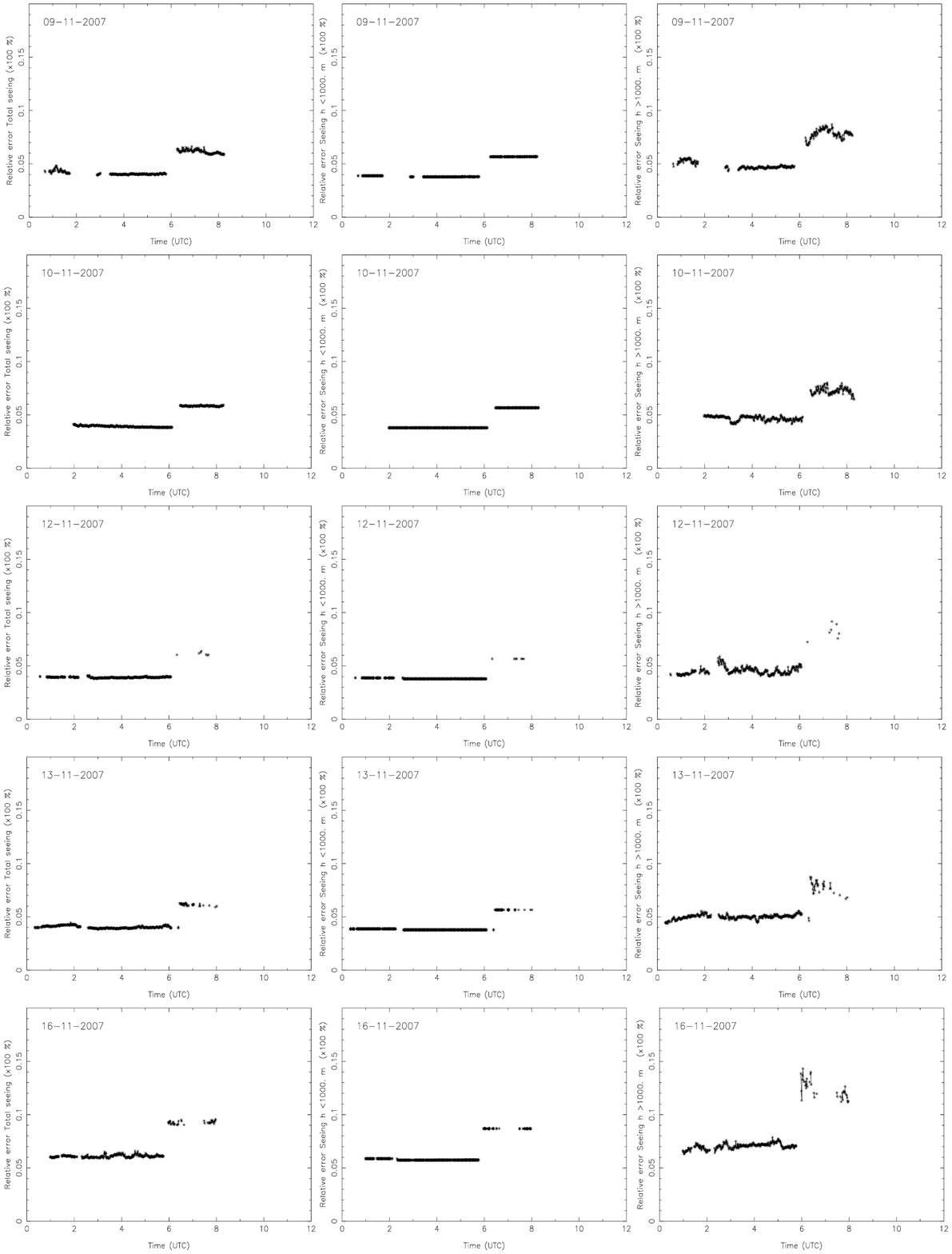}
\end{center}
\caption{Temporal evolution of the relative error of the total seeing $\varepsilon_{TOT}$ (left), the boundary layer seeing $\varepsilon_{BL}$ (centre) and the seeing in the free atmosphere $\varepsilon_{FA}$ (right) during the 20 nights. $\varepsilon_{BL}$  is defined as the seeing integrated for h $<$ 1km, $\varepsilon_{FA}$ as the seeing for h $>$ 1km. The dates are in local time (LT). Ex: 9/11/2007 corresponds to 10/11/2007 in UT. Y-axis values have to be multiplied by 100 and expressed in (\%). Ex: 0.1 means 10 \%.}
\label{rel_err_see_page1}
\end{figure*}

From Fig.\ref{rel_err_see_page1} up to Fig.\ref{rel_err_see_page4} are shown the temporal evolution of the relative error of the total seeing, the boundary layer ($h$ $<$ 1000~m) and the free atmosphere ($h$ $>$ 1000~m) contributions for all the 20 nights of the PAR2007 site testing campaign.

%
%

\begin{figure*}
\begin{center}
\includegraphics[width=15cm]{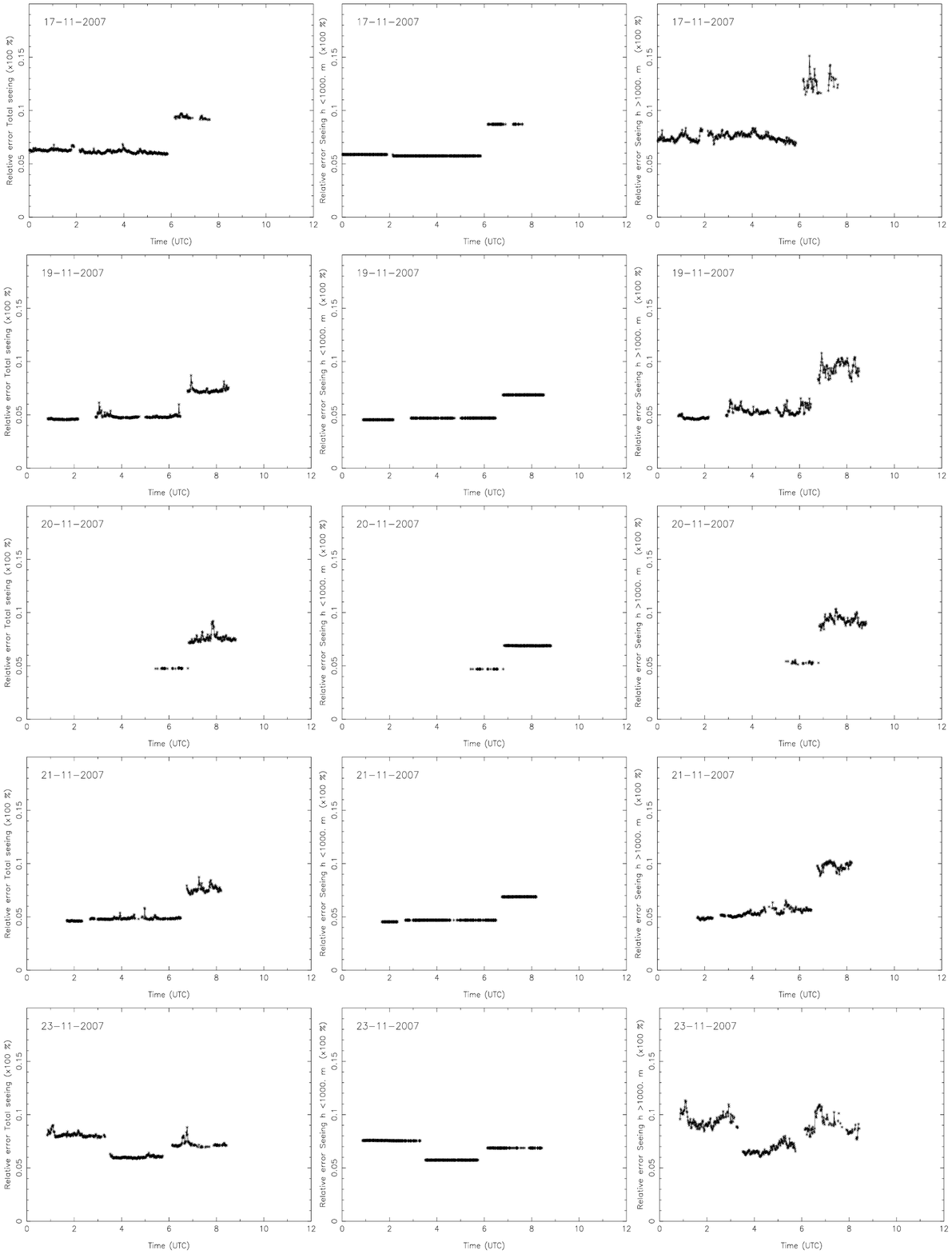}
\end{center}
\caption{It follows Fig. \ref{rel_err_see_page1}.}
\label{rel_err_see_page2}
\end{figure*}

\begin{figure*}
\begin{center}
\includegraphics[width=15cm]{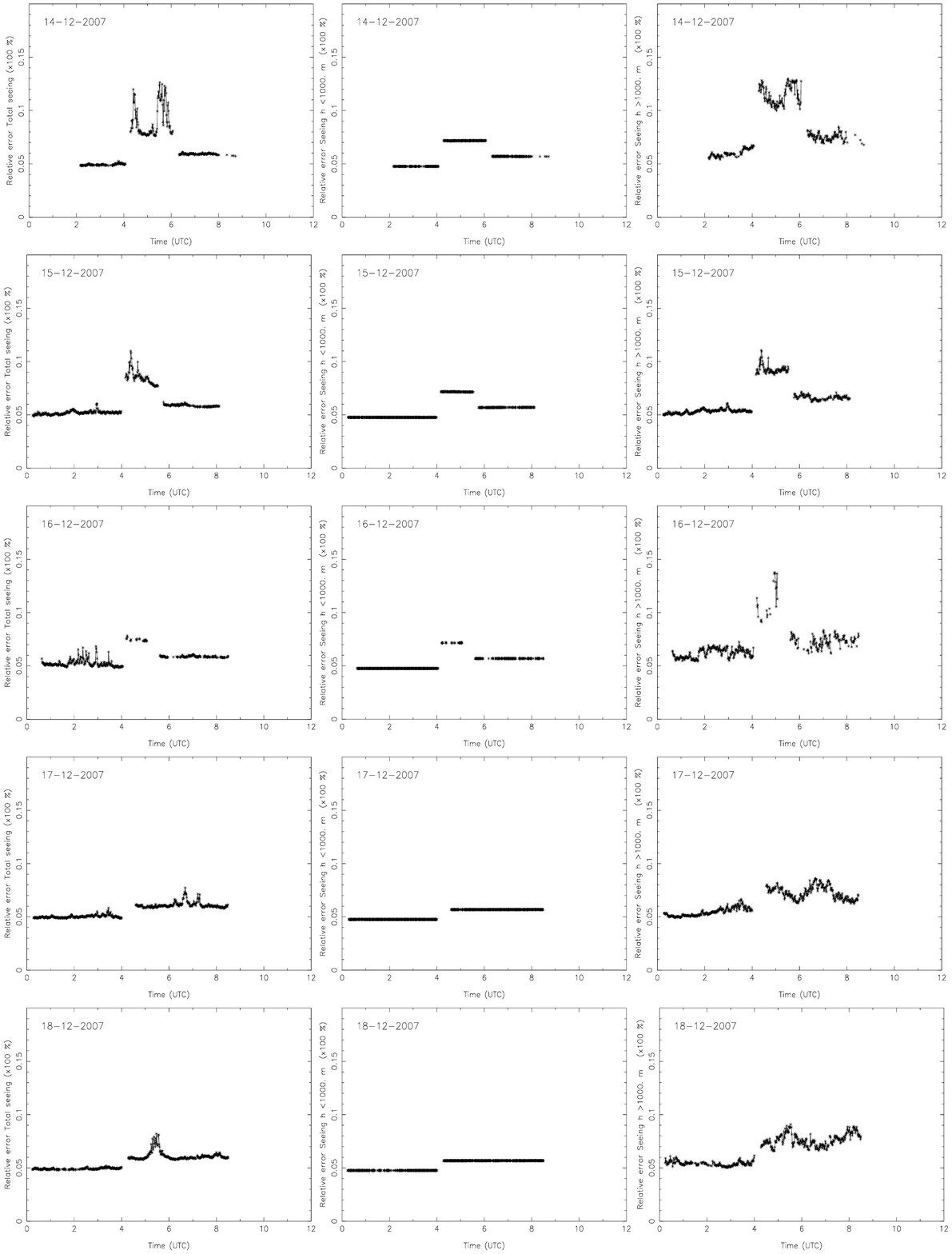}
\end{center}
\caption{It follows Fig. \ref{rel_err_see_page1}.}
\label{rel_err_see_page3}
\end{figure*}

\begin{figure*}
\begin{center}
\includegraphics[width=15cm]{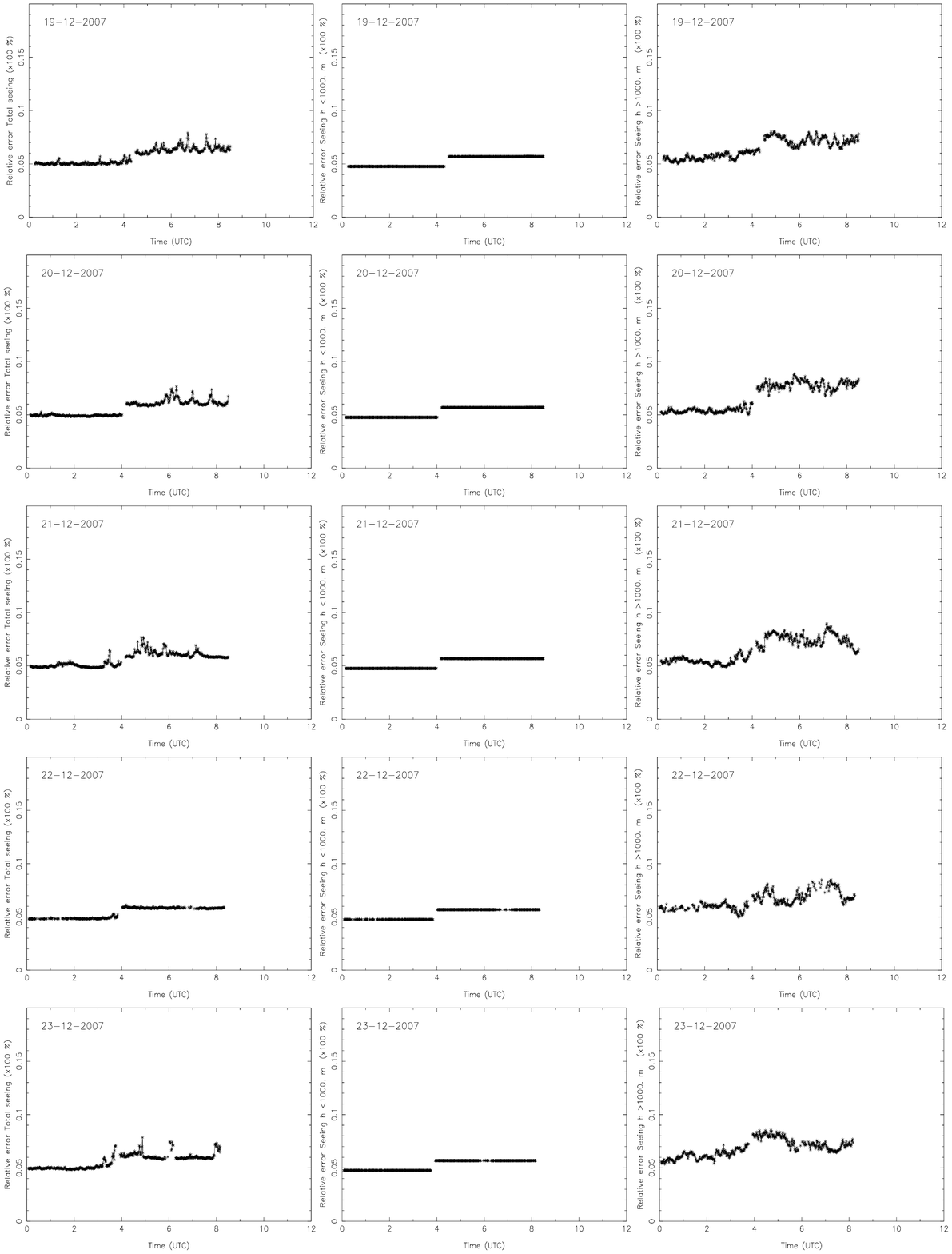}
\end{center}
\caption{It follows Fig. \ref{rel_err_see_page1}.}
\label{rel_err_see_page4}
\end{figure*}

\section{Temporal evolution of the relative error of the isoplanatic angle for the individual nights}
\label{annex_iso}

\begin{figure*}
\begin{center}
\includegraphics[width=15cm]{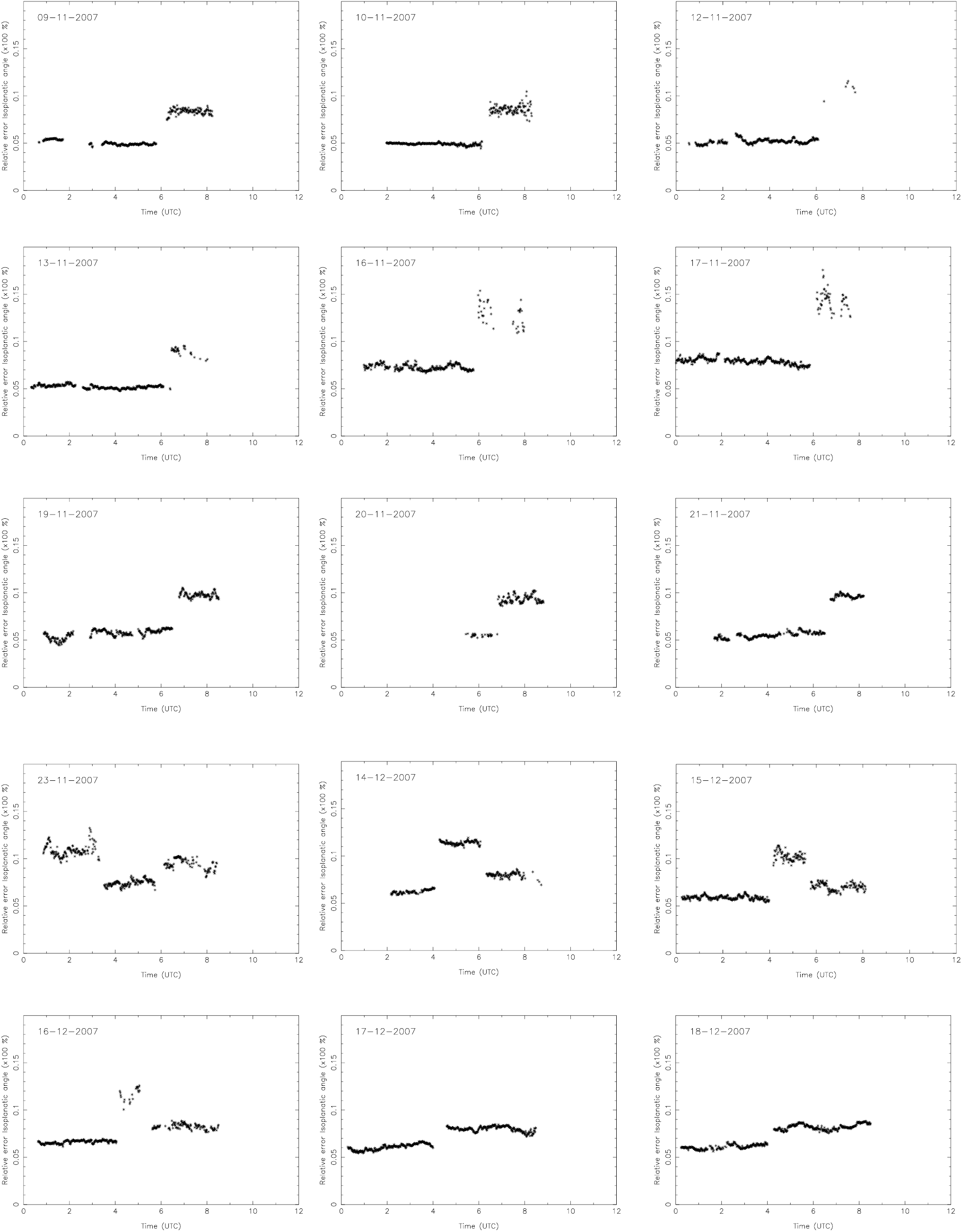}
\end{center}
\caption{Temporal evolution of the relative error of the isoplanatic angle $\theta_{0}$ during the 20 nights of the site testing campaign of November/December 2007. The dates are in local time (LT). Ex: 9/11/2007 corresponds to 10/11/2007 in UT. Y-axis values have to be multiplied by 100 and expressed in (\%). Ex: 0.1 means 10 \%.}
\label{iso_pag1}
\end{figure*}

Fig.\ref{iso_pag1} shows the temporal evolution of the relative error of the isoplanatic angle for all the 20 nights of the PAR2007 site testing campaign.

%
%

\begin{figure*}
\begin{center}
\includegraphics[width=15cm]{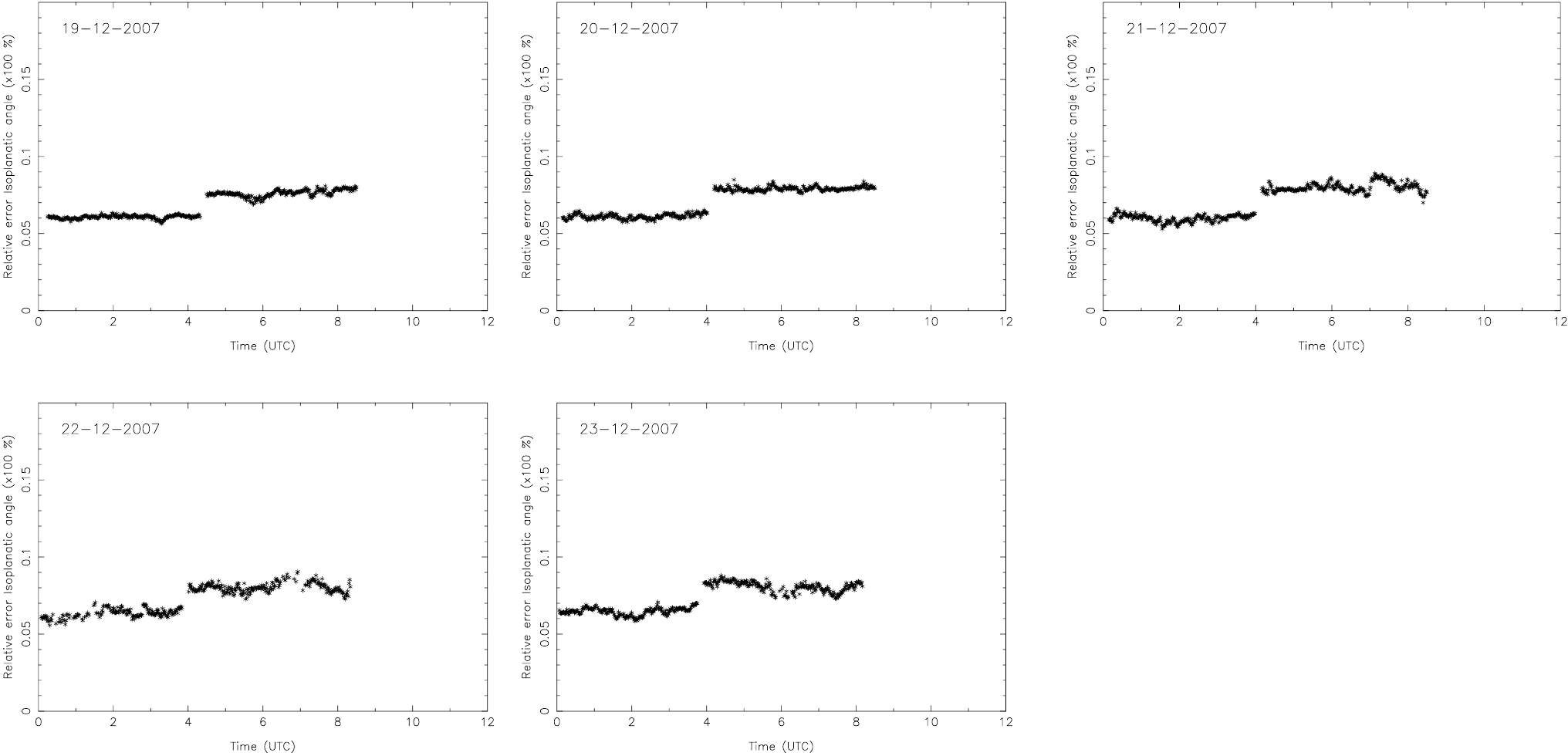}
\end{center}
\caption{It follows Fig. \ref{iso_pag1}.}
\label{iso_pag2}
\end{figure*}

\section{Temporal evolution of the seeing for the individual nights}
\label{see_temp_evol}

\begin{figure*}
\begin{center}
\includegraphics[width=15cm]{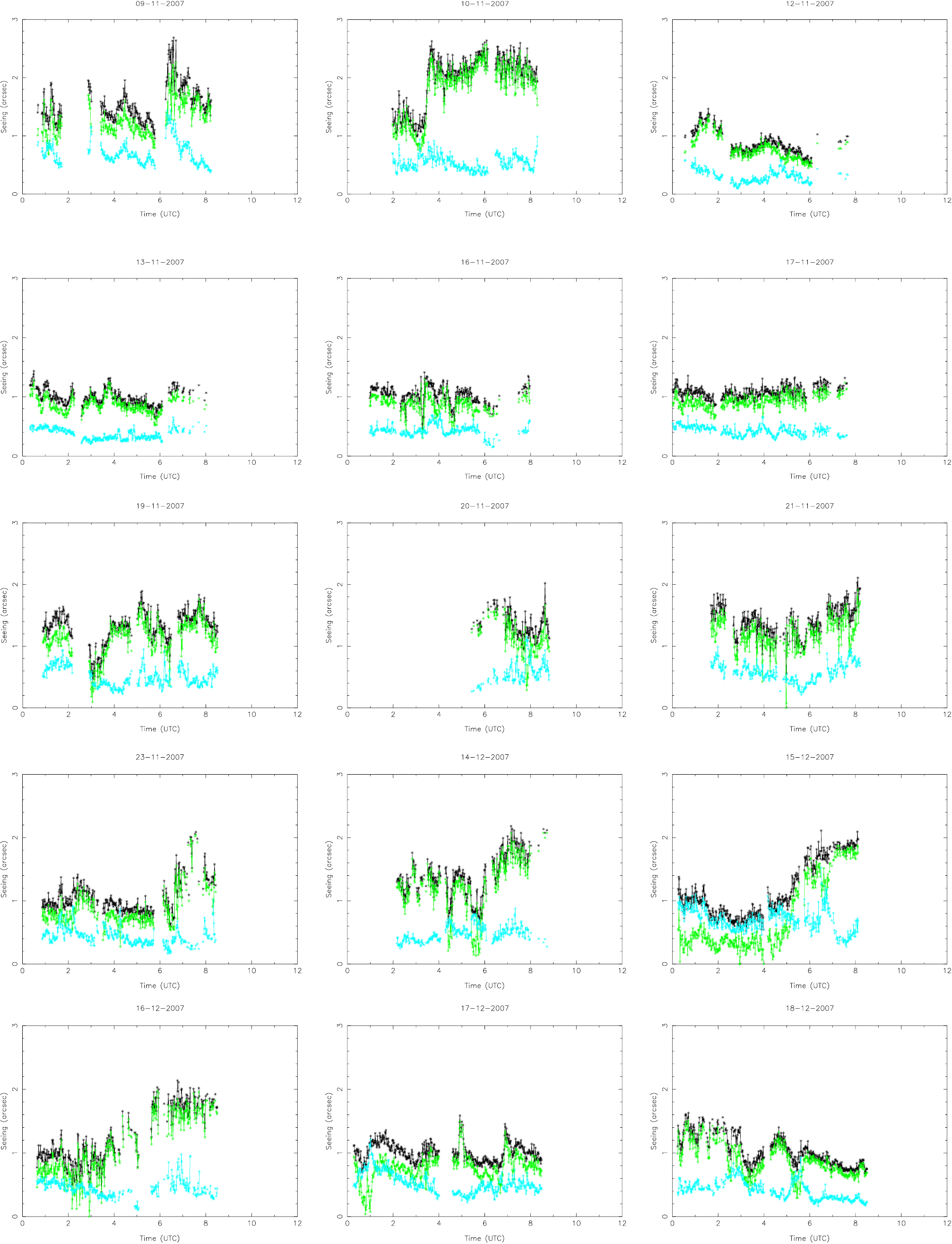}
\end{center}
\caption{Temporal evolution of the seeing during the 20 nights of the site testing campaign of November/December 2007. The dates are in local time (LT). Ex: 9/11/2007 corresponds to 10/11/2007 in UT. Black line: total seeing. Green line: seeing in the boundary layer ($h$ $<$ 1000~m). Light blue line: seeing in the free atmosphere ($h$ $>$ 1000~m). }
\label{see_temp_evol_1}
\end{figure*}

\begin{figure*}
\begin{center}
\includegraphics[width=15cm]{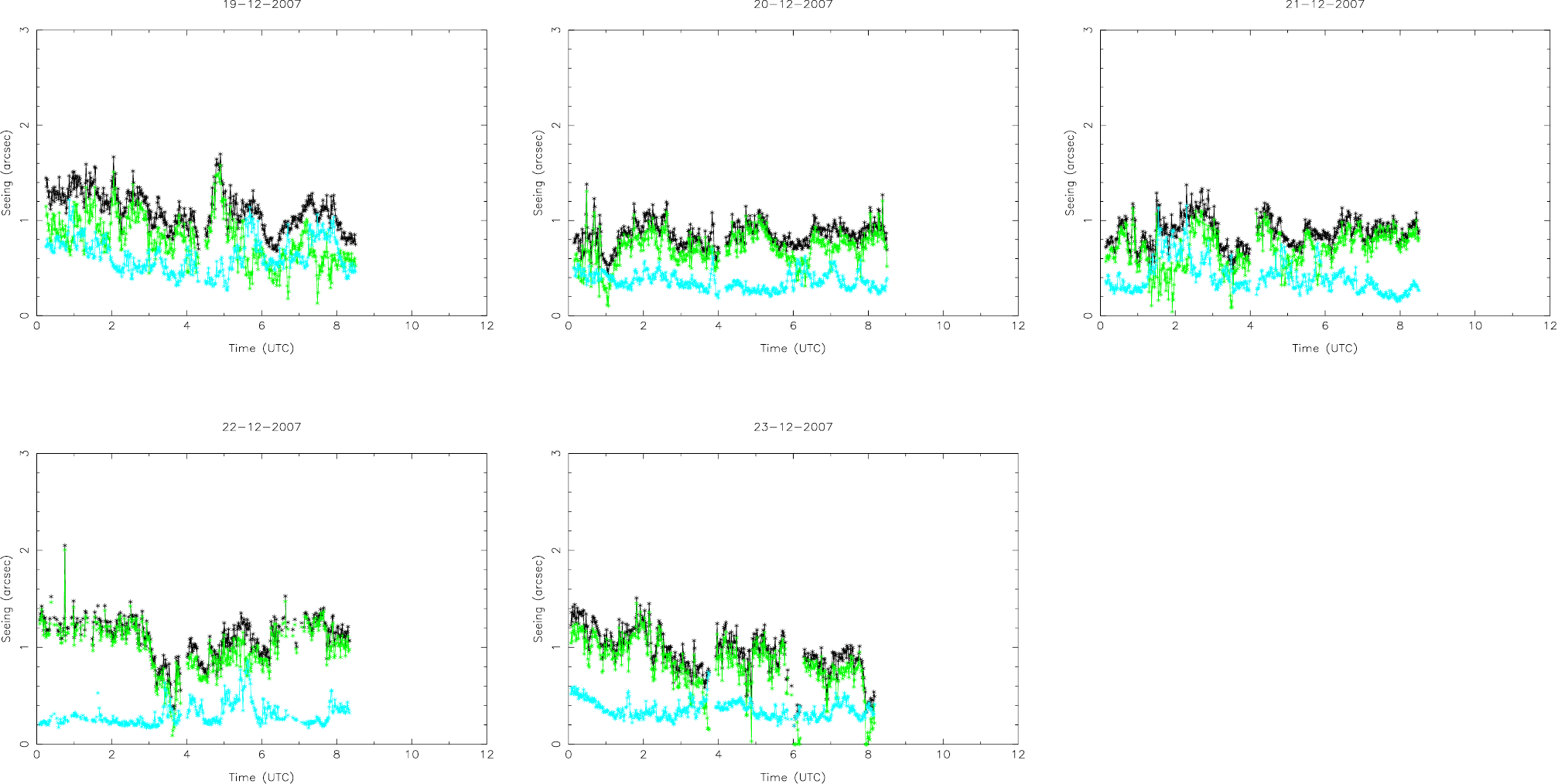}
\end{center}
\caption{It follows Fig. \ref{see_temp_evol_1}.}
\label{see_temp_evol_2}
\end{figure*}

Fig.\ref{see_temp_evol_1} and Fig.\ref{see_temp_evol_2} show the temporal evolution of the seeing (total, boundary layer and free atmosphere) for all the 20 nights of the PAR2007 site testing campaign.

%
%

\section{Temporal evolution of the isoplanatic angle for the individual nights}
\label{iso_temp_evol}

\begin{figure*}
\begin{center}
\includegraphics[width=15cm]{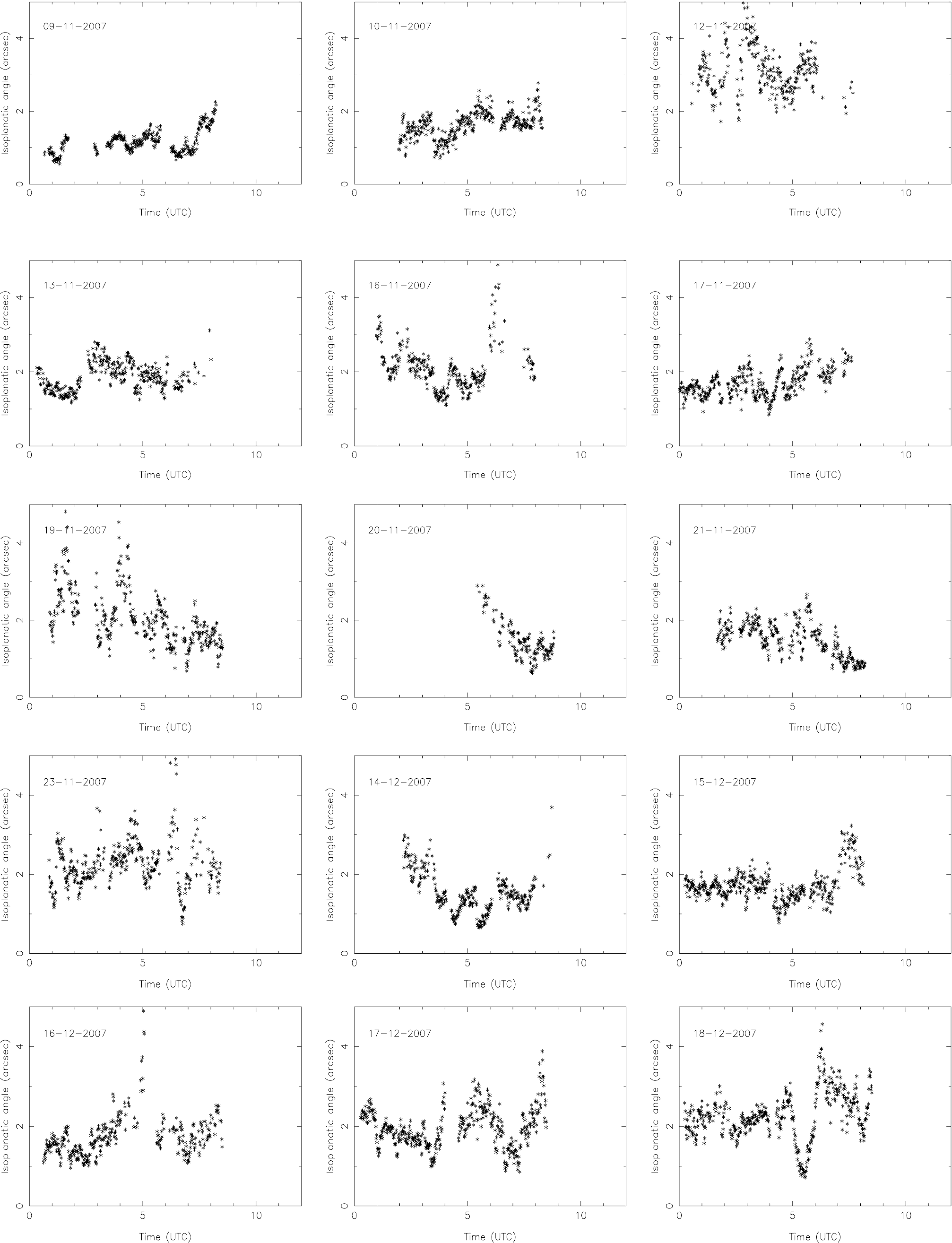}
\end{center}
\caption{Temporal evolution of the isoplanatic angle during the 20 nights of the site testing campaign of November/December 2007. The dates are in local time (LT). Ex: 9/11/2007 corresponds to 10/11/2007 in UT. The integral is calculated up to 20km from the ground. See text for discussion. }
\label{iso_temp_evol_1}
\end{figure*}

\begin{figure*}
\begin{center}
\includegraphics[width=15cm]{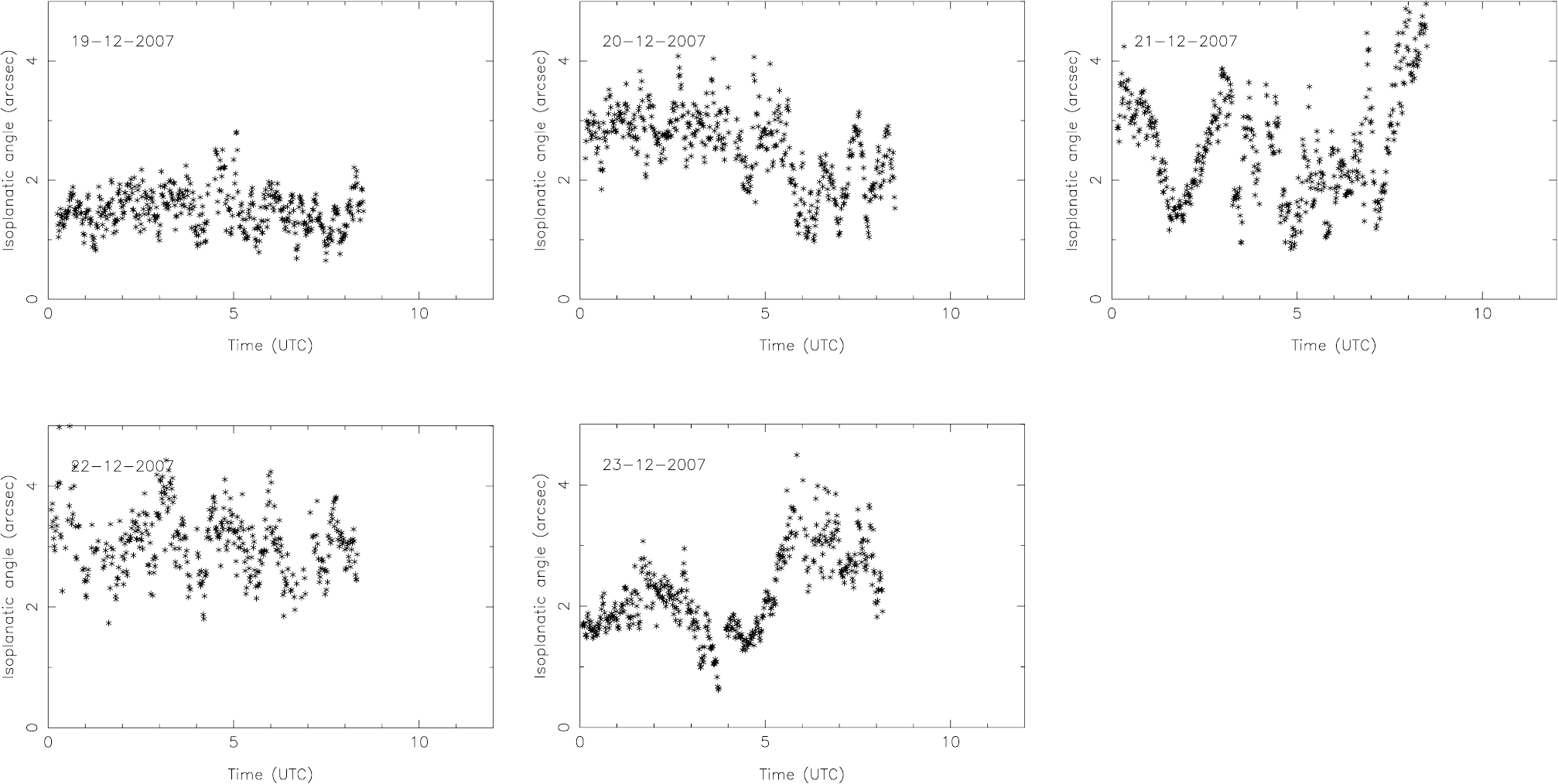}
\end{center}
\caption{It follows Fig. \ref{iso_temp_evol_1}.}
\label{iso_temp_evol_2}
\end{figure*}

Fig.\ref{iso_temp_evol_1} and Fig.\ref{iso_temp_evol_2} show the temporal evolution of the isoplanatic angle for all the 20 nights of the PAR2007 site testing campaign.

%
%

\label{lastpage}
\end{document}